\documentclass[12pt]{iopart}

\usepackage{graphicx}
\usepackage{dcolumn}
\usepackage{amssymb,epsfig}
\expandafter\let\csname equation*\endcsname\relax
\expandafter\let\csname endequation*\endcsname\relax
\usepackage{amsmath}
\usepackage{paralist}
\usepackage{comment}
\usepackage{multirow}
\usepackage{color,soul}
\usepackage{suffix}
\usepackage{wasysym}
\usepackage[normalem]{ulem}
\usepackage{stackrel}

\allowdisplaybreaks

\begin{document}

\title[Diffraction of light by plasma in the solar system]{Diffraction of light by plasma in the solar system}

\author{Slava G. Turyshev$^{1}$, Viktor T. Toth$^2$}

\address{$^1$Jet Propulsion Laboratory, California Institute of Technology,\\
4800 Oak Grove Drive, Pasadena, CA 91109-0899, USA \\ [10pt]
$^2$Ottawa, Ontario K1N 9H5, Canada}
\vspace{10pt}
\begin{indented}
\item[]January 2019
\end{indented}

\begin{abstract}
We study the propagation of electromagnetic (EM) waves in the solar system and develop a Mie theory that accounts for the refractive properties of the free electron plasma in the extended solar corona. We use a generic model for the electron number density distribution and apply the eikonal approximation to find a solution in terms of Debye potentials, which is then used to determine the EM field both within the inner solar system and at large heliocentric distances. As expected, the solution for the EM wave propagating  through the solar  system is characterized by a plasma-induced phase shift and related change in the light ray's direction of propagation. Our approach quantitatively accounts for these effects, providing a wave-optical treatment for diffraction in the solar plasma. As such, it may be used in practical applications involving big apertures, large interferometric baselines or otherwise widely distributed high-precision astronomical instruments.

\end{abstract}

%
%
%
%
%

\section{Introduction}

The propagation of electromagnetic (EM) signals in a refractive medium is highly dependent on the frequency of the wave and the properties of the medium \cite{Ginzburg-book-1964,Landau-Lifshitz:1979}. The solar plasma is such a refractive medium, influencing astronomical observations conducted in the solar system. With the advent of solar system exploration by space probes, the influence of plasma on interplanetary radio communication links was studied extensively \cite{Ginzburg-Zheleznyakov:1959,Muhleman-etal:1977,Tyler-etal:1977} and it is well characterized \cite{Giampieri:1994kj,Bertotti-Giampieri:1998,DSN-handbook-2017,Verma-etal:2013}. Similar efforts were conducted to account for the solar plasma in astronomical observations conducted at optical and IR wavelengths as well as $\gamma$-rays and X-rays (see \cite{Huber-etal:2013} and references therein).

Many astrophysical phenomena require precision observations, in which the solar plasma contributes significantly. The effects of the solar plasma are especially prominent at radio wavelengths. It is necessary to account for these effects in diverse applications such as very long baseline interferometry (VLBI) and spacecraft navigation. These effects are also  relevant to experiments aiming to achieve very high magnification via gravitational lensing \cite{Clegg:1997ya}. One example is the solar gravitational lens (SGL), where bending of light by the gravitational field of the Sun is used to achieve extreme light amplification and angular resolution \cite{Turyshev-Toth:2017}, which, in the foreseeable future, may become a means to obtain direct megapixel scale imaging and spectroscopy of Earth-like planets orbiting nearby stars \cite{Turyshev-Toth:2017,Turyshev-etal:2018}. Thus, an appropriate description of light propagation in the refractive medium of the solar plasma is an important problem.

Propagation of light with wavelength $\lambda$ in the vicinity of a large sphere with radius when $R\gg \lambda$ is typically described using the geometric optics approximation. With this approach, we can study trajectories of individual light rays and describe the plasma-induced phase shift and related frequency change  (e.g., \cite{Muhleman-etal:1977,Tyler-etal:1977,Bertotti-Giampieri:1998,DSN-handbook-2017,Verma-etal:2013}). However, for modern high-precision astronomical observations a wave-theoretical treatment may be preferable. Although it is known, the Mie theory \cite{Mie:1908,Born-Wolf:1999} provides a rather good framework to develop such a treatment, however, no such developments are known to describe the situation of a very large opaque sphere surrounded by plasma.

Recently \cite{Turyshev-Toth:2018}, we presented a wave-optical description of the shadow cast by a large opaque sphere. Specifically, we considered the scattering of EM waves by the large sphere and developed a Mie theory that accounts for the presence of an obscuration. We were able to determine that there is no EM field in the shadow in the wave zone behind the sphere besides that related to the Poisson-Arago bright spot. In the present study, we rely on the tools and methods developed in \cite{Turyshev-Toth:2018} to describe light propagation in the vicinity of the Sun. Our main concern is the effect due to the dispersive nature of the solar plasma on the EM field as it propagates through the solar system.

This paper is organized as follows. In Section \ref{sec:wave-plasma} we discuss the solar corona, modeled as a free electron nonmagnetic plasma. For this, we use the most generic plasma model and develop a solution for Maxwell's equations, characterizing the EM field in such a refractive medium. In Section~\ref{sec:Debye-pot} we derive a solution of the EM field equations in terms of Debye potentials. In Section~\ref{sec:sol-EM-Deb} we introduce the eikonal approximation, to deal with the long-range component of the scattering potential. We account for the short-range potential and obtain a full solution for the EM field. In Section~\ref{sec:diff-large-sph} we complete the solution by setting up appropriate boundary conditions for the EM field and determine this field in all regions of the solar system. Finally, in Section~\ref{sec:disc} we discuss results and practical applications.

\section{The extended solar corona}
\label{sec:eqs-g+p}
\label{sec:wave-plasma}

To describe the propagation of an EM wave in the solar system, we first need to introduce a model for the solar plasma and the interplanetary medium that would cover the heliocentric ranges of interest. Thermonuclear reactions occurring inside the Sun result in the emission of large amounts of energy  \cite{Lang-book:2009}. Much of this energy is released in the form of EM radiation.
However, the Sun also emits a stream of charged particles, known as the solar wind. The solar wind is ionized: electrons and protons are separate, yielding a gaseous medium in which electrons are free, with no restoring force due to nearby atomic nuclei. This plasma extends to the outer solar system.

For an EM wave of angular frequency $\omega$ propagating through plasma, the dielectric permittivity of the plasma, in general, is defined as \cite{Landau-Lifshitz:1979}:
\begin{equation}
\hskip -20pt
 \epsilon (t,{\vec r})=1- \frac{4\pi n_e (t,{\vec r})e^2 }{m_e\omega^2}= 1 - \frac{\omega_p^{ 2}}{\omega^2}, \qquad {\rm where} \qquad
 \omega^2_{\tt p}=\frac{4\pi n_ee^2}{m_e},
 \label{eq:eps}
 \end{equation}
where $e$ is the electron's charge, $m_e$ is its mass and $n_e=n_e(t,{\vec r})$ is the electron number density. The quantity $\omega_p$ is known as the electron plasma (or Langmuir) frequency. It is reasonable to assume that the solar plasma is nonmagnetic, i.e., its magnetic permeability is $\mu=1$.

Therefore, in order to evaluate the plasma contribution to Maxwell's equations, we need to know the electron number density along the path. In general, of course, the electron plasma density shows temporal variability. Thus, we start by decomposing the electron number density $n_e$ into a steady-state term part $\overline{n}_e({\vec r})$ plus a temporal fluctuation  $\delta n_e(t, {\vec r})$:
\begin{equation}
n_e(t, {\bf r})= \overline{n}_e({\vec r})+ \delta n_e(t, {\vec r}).
\label{eqelcont}
\end{equation}

The variability of the solar atmosphere has no preferred time scale. Variations in the electron number density, $ \delta n_e(t, {\bf r})$, can be of a magnitude equal to that of the steady-state term, $\overline{n}_e({\vec r})$, \cite{Armstrong-etal:1979}. These variations are carried along by the solar wind, at a typical speed of $\sim 400$~km/s; over integration times of $\sim 10^3$ seconds, the spatial scale of the fluctuations will therefore be comparable to the solar radius.
As these deviations are unpredictable in nature, their contributions must be treated as noise \cite{Turyshev-Toth:2018-grav-plasma}.

In contrast, the steady-state component of the solar corona is well understood, and the magnitude of its contribution can be estimated. In fact, much of our knowledge about the solar plasma comes from the tracking of spacecraft in the inner solar system \cite{Muhleman-etal:1977,Tyler-etal:1977,Giampieri:1994kj,Bertotti-Giampieri:1998,Verma-etal:2013,Lang-book:2009,Muhleman-Anderson:1981,Lang-ebook:2010}. Distant spacecraft provide information about the extent of solar plasma as we approach interstellar space \cite{Belcher-etal:1993,Stone-etal:1996,Decker-etal:2005}. The heliopause at $\sim 130$~AU is the last frontier of the heliosphere, the region of space dominated by the solar wind. At this distance, the momentum density of the solar wind is no longer sufficient to repulse the rarefied hydrogen and helium that is found in interstellar space. The region just inside the heliopause is called the heliosheath: the turbulent region where the solar wind is slowed and compressed by interstellar pressure. The inner boundary of the heliosheath, the termination shock, represents the region where the solar wind first collides with the interstellar medium.

As a result, in what follows, to describe the plasma distribution throughout the entire solar system, we assume that the electron number density in the solar corona and the solar wind is steady-state, spherically symmetric\footnote{Although they may be incorporated in the model (\ref{eq:n-eps_n-ism}), we ignore any corrections that depend on the heliographic latitude.} and may be parameterized in the following generic form:
{}
 \begin{eqnarray}
\overline{n}_e({\vec r})=
\left\{
                \begin{array}{lll}
~~0,& \hskip 20pt \phantom{R_\odot}0\leq r<R_\odot,\\
\sum_i \alpha_i \Big(\dfrac{R_\odot}{r}\Big)^{\beta_i},\quad&
\hskip 20pt \phantom{0}R_\odot\leq r\leq R_\star,\\
~n_0,&\hskip 20pt \phantom{R_\odot 0\leq{}}r>R_\star,
 \end{array}
              \right.
 \label{eq:n-eps_n-ism}
 \end{eqnarray}
where $\alpha_i$ and $\beta_i$ are empirically determined values and $R_\odot$ is the solar radius. Note that $i\geq2$, which is needed to replicate the $1/r^2$ behavior of the solar wind at large distances from the Sun, where $r\gg R_\odot$. The value $R_\star$ represents the heliocentric distance to the termination shock\footnote{In addition to being physically justified, the model (\ref{eq:n-eps_n-ism}) also has the mathematical advantage as it helps to avoid divergences when solving the differential equations for the EM field.  The primary concern is, of course, the $1/r^2$ term that, as it is well known  \cite{Friedrich-book-2006,Friedrich-book-2013}, leads to divergences when integrating to infinity (as would be in the case when investigating plane waves incoming from infinity), forcing the introduction of cut-offs. As we shall see later, our model is self-consistent both physically and mathematically, leading to finite results in all regions of interest, with no significant dependance on the choice of $R_\star$.}, which we take to be at $R_\star\simeq100$~AU, roughly corresponding to the distance to the inner boundary of the heliosphere. The symbol $n_0$ represents the electron number density in the interstellar medium. The presence of this term in the model is for completeness only as it does not diffract light due to its assumed uniform and homogenous behavior. Specifically, for distances $r>R_\star$, trajectories of light rays do not change direction as their phase is uniformly delayed due to the uniform background given by $n_0$ in (\ref{eq:n-eps_n-ism}). Therefore, without loss of generality, we can take $n_0=0$ or reinstate its nonzero value if needed. Note that $\overline{n}_e({\vec r})$ is not continuous at $R_\star$, reflecting the abrupt change in plasma density at the termination shock as reported by the Voyager 1 spacecraft \cite{Burlaga-etal:2005}. As a light ray crosses the termination shock at $R_\star$ and proceeds into the inner solar system, it is now affected by the second term in (\ref{eq:n-eps_n-ism}). For the range of heliocentric distances $R_\odot\leq r\leq R_\star$ light rays are refracted by the solar plasma with their phase being delayed and their trajectories bent.  Finally, as light reaches the surface of the Sun at $r=R_\odot$, it is absorbed by the Sun resulting in a geometric shadow behind the Sun \cite{Turyshev-Toth:2018}.

Expressions (\ref{eq:eps})--(\ref{eq:n-eps_n-ism}) represent what we call the {\it extended solar corona} model introduced within the entire solar system and extending beyond the termination shock. The solar plasma modeled by (\ref{eq:n-eps_n-ism}) has a variable, negative index of refraction. As such, it has the effect of deflecting outwards the wavefronts of light passing by the Sun.

The steady-state behavior of the solar plasma is known reasonably well. There are several plasma models found in the literature that we can utilize (see discussion in \cite{Turyshev-Andersson:2002}). One specific example of the model (\ref{eq:n-eps_n-ism}) with particular values for $\alpha_i$ and $\beta_i$ is
\cite{Tyler-etal:1977,Turyshev-Andersson:2002}:
{}
\begin{equation}
\hskip -20pt
\overline{n}_e(r)= \Big[2.99\times10^8 \,\Big(\frac{R_\odot}{r}\Big)^{16} +
1.55\times10^8 \, \Big(\frac{R_\odot}{r}\Big)^{6}+ {3.44\times 10^5}\, \Big(\frac{R_\odot}{r}\Big)^{2}\Big]~{\rm
cm}^{-3},
\label{eq:model}
\end{equation}
where $r\geq R_\odot$. The coefficients in this model are determined empirically by processing the tracking data using radio communication links to interplanetary spacecraft. In fact, the model given by Eq.~(\ref{eq:model}) was used to process Cassini tracking data \cite{Anderson-etal:2002,Turyshev-Toth:2010LRR}. While other models exist \cite{Verma-etal:2013}, they are generally compatible with the Cassini model (\ref{eq:model}).

Evaluating (\ref{eq:model}) for rays of light passing near the Sun with the smallest impact parameter, $b=R_\odot$, we see that the electron number density, at most, would be of the order of $\overline{n}_e(r)\lesssim 10^{10} ~{\rm cm}^{-3}$, which implies a frequency of $\nu_p=\omega_p/2\pi=\sqrt{n_ee^2/\pi m_e}\lesssim 1$~GHz. For optical frequencies ($\nu=c/\lambda\sim 300$~THz) and at the smallest impact parameter, (\ref{eq:eps}) contributes at most to the order of $(\omega_p/\omega)^2\lesssim10^{-11}$, even though for radio frequencies ($\nu\sim 10$~GHz) this ratio is much higher: $(\omega_p/\omega)^2\lesssim10^{-2}$. As the main subject of our interest is visible or near-IR light, we only consider terms that are linear with respect to the plasma contribution and omit higher order terms.

The plasma frequency $ \omega_{\tt p}^2$ in Eq.~(\ref{eq:eps}), in the case of the spherically symmetric plasma distribution
(\ref{eq:n-eps_n-ism}), in the range of heliocentric distances,  $R_\odot\leq r\leq R_\star$, has the form
 \begin{equation}
 \omega_{\tt p}^2=\frac{4\pi e^2}{m_e}
\sum_i \alpha_i\Big(\frac{R_\odot}{r}\Big)^{\beta_i}.
 \label{eq:n_n-ss}
 \end{equation}
This model for the plasma frequency in the extended solar corona allows us to study the influence of solar plasma on the propagation of EM waves throughout the solar system in the range of heliocentric distances given by $R_\odot\leq r\leq R_\star$.

\section{The EM field and Debye potentials}
\label{sec:Debye-pot}

We begin our derivation of the EM field equations in the presence of plasma with presenting Maxwell's source-free field equations in their well-known form \cite{Born-Wolf:1999}:
{}
\begin{eqnarray}
\hskip -50pt
{\rm curl}\,{\vec E}&=&-  \frac{1}{c}\frac{\partial \,(\mu{\vec H})}{\partial t},
\quad
{\rm curl}\,{\vec H}= \frac{1}{c}\frac{\partial \,(\epsilon{\vec E})}{\partial t},
\quad
~{\rm div}\big(\epsilon {\vec E}\big)\,=\,0,
\label{eq:rotE_fl}
\quad
{\rm div }\big(\mu {\vec B}\big)\,=\,0.
\label{eq:rotH_fl}
\end{eqnarray}
Equations~(\ref{eq:rotE_fl}) capture the contribution of the solar plasma to the propagation of light in the vicinity of the Sun. Following closely the derivation presented in \cite{Turyshev-Toth:2017}, we now consider a solution to these equations. Assuming, as usual \cite{Born-Wolf:1999}, the time dependence $\exp(-i\omega t)$ and taking\footnote{When an EM wave is propagating in an electron plasma, its frequency is given by the dispersion relation $\omega^2(k)=k^2 c^2+\omega_{\tt p}^2(k)$ \cite{Landau-Lifshitz:1979}. That is, the plasma modifies the dispersion relation and affects the group and phase velocities. Realizing that the electron number density for the solar plasma is at most $\overline{n}_e(r)\lesssim 6\times 10^{8}~{\rm cm}^{-3}$ \cite{Verma-etal:2013,Mercier-Chambe:2015}, using (\ref{eq:eps}), we compute the largest relevant value of $\omega_{\tt p}^2(k)$ that yeilds $\omega^2(k)=k^2c^2\big(1+ 5.38\times 10^{-13}(\lambda/1~\mu{\rm m})^2\big)$. Therefore, throughout this paper we use $\omega^2=k^2 c^2\big(1+{\cal O}(10^{-12})\big)$, signifying that at the optical and near-IR wavelengths relevant to the SGL, $\lambda\simeq 1~\mu$m, the difference between the group and phase velocities can be neglected.}  $k=\omega/c$, the time-independent parts of the electric and magnetic vectors must satisfy Maxwell's equations (\ref{eq:rotE_fl})
in their time-independent form:
{}
\begin{eqnarray}
{\rm curl}\,{\vec { E}}&=&ik\mu \,{\vec { H}}, \qquad
\label{eq:rotH_fl**=}
{\rm curl}\,{\vec { H}}=-ik\epsilon \,{\vec {  E}}.
\label{eq:rotE_fl**=}
\end{eqnarray}

In the case of a static, spherically symmetric plasma distribution, solving (\ref{eq:rotH_fl**=}) is most straightforward. Following \cite{Born-Wolf:1999,Turyshev-Toth:2017}, we obtain the complete solution of (\ref{eq:rotH_fl**=}) in terms of the electric and magnetic Debye potentials,
${}^e\Pi$ and ${}^m\Pi$:
{}
\begin{eqnarray}
{ { E}}_r&=&
\frac{1}{\sqrt{\epsilon}}\Big\{\frac{\partial^2 }{\partial r^2}
\Big[\frac{r\,{}^e{\hskip -1pt}\Pi}{\sqrt{\epsilon}}\Big]+\Big(\epsilon\mu\,k^2 -\sqrt{\epsilon}\big(\frac{1}{\sqrt{\epsilon}}\big)''\Big)\Big[\frac{r\,{}^e{\hskip -1pt}\Pi}{\sqrt{\epsilon}}\Big]\Big\},
\label{eq:Dr-em0}\\[3pt]
{ {  E}}_\theta&=&
\frac{1}{\epsilon r}\frac{\partial^2 \big(r\,{}^e{\hskip -1pt}\Pi\big)}{\partial r\partial \theta}+\frac{ik}{r\sin\theta}
\frac{\partial\big(r\,{}^m{\hskip -1pt}\Pi\big)}{\partial \phi},
\label{eq:Dt-em0}\\[3pt]
{ {  E}}_\phi&=&
\frac{1}{\epsilon r\sin\theta}
\frac{\partial^2 \big(r\,{}^e{\hskip -1pt}\Pi\big)}{\partial r\partial \phi}-\frac{ik}{r}
\frac{\partial\big(r\,{}^m{\hskip -1pt}\Pi\big)}{\partial \theta},
\label{eq:Dp-em0}\\
{ {  H}}_r &=&
\frac{1}{\sqrt{\mu}}\Big\{\frac{\partial^2}{\partial r^2}\Big[\frac{r\,{}^m{\hskip -1pt}\Pi}{\sqrt{\mu}}\Big]+\Big(\epsilon\mu\,k^2-\sqrt{\mu}\big(\frac{1}{\sqrt{\mu}}\big)''\Big)\Big[\frac{r\,{}^m{\hskip -1pt}\Pi}{\sqrt{\mu}}\Big]\Big\},
\label{eq:Br-em0}\\[3pt]
{ {  H}}_\theta&=&
-\frac{ik}{r\sin\theta} \frac{\partial\big(r\,{}^e{\hskip -1pt}\Pi\big)}{\partial \phi}+\frac{1}{\mu r}
\frac{\partial^2 \big(r\,{}^m{\hskip -1pt}\Pi\big)}{\partial r\partial \theta},
\label{eq:Bt-em0}\\[3pt]
{ { H}}_\phi&=&
\frac{ik}{r}\frac{\partial\big(r\,{}^e{\hskip -1pt}\Pi\big)}{\partial \theta}+\frac{1}{\mu r\sin\theta}
\frac{\partial^2 \big(r\,{}^m{\hskip -1pt}\Pi\big)}{\partial r\partial \phi},
\label{eq:Bp-em0}
\end{eqnarray}
where the potentials ${}^e{\hskip -1pt}\Pi$ and  ${}^m{\hskip -1pt}\Pi$ satisfy the following wave equations:
\begin{eqnarray}
\hskip -50pt
\Big(\Delta+\epsilon\mu\,k^2-\sqrt{\epsilon}\big(\frac{1}{\sqrt{\epsilon}}\big)''\Big)\Big[\frac{\,{}^e{\hskip -1pt}\Pi}{\sqrt{\epsilon}}\Big]=0,
\label{eq:Pi-eq+wew1*+}
\qquad
\Big(\Delta+\epsilon\mu\,k^2-\sqrt{\mu}\big(\frac{1}{\sqrt{\mu}}\big)''\Big)\Big[\frac{\,{}^m{\hskip -1pt}\Pi}{\sqrt{\mu}}\Big]=0.
\label{eq:Pi-eq+wmw1*}
\end{eqnarray}

In the case of the weakly interacting, spherically symmetric free electron plasma of the extended solar corona, Eqs.~(\ref{eq:Pi-eq+wew1*+}) may be simplified.  First of all, using (\ref{eq:eps}) together with (\ref{eq:n-eps_n-ism}) for $\epsilon$, while setting $\mu=1$, we can rewrite the left equation in (\ref{eq:Pi-eq+wew1*+}) as the equation that describes scattering in the presence of the plasma:
{}
\begin{eqnarray}
\Big\{\Delta +k^2\big(1-\frac{\omega_p^{ 2}(r)}{\omega^2}\big)-\frac{(\omega_{\tt p}^2)''}{4\omega^2}\Big\}\Big[\frac{\,{}^e{\hskip -1pt}\Pi}{\sqrt{\epsilon}}\Big]=0.
\label{eq:Pi-eq*0+*}
\end{eqnarray}

A similar equation (but without the last, fourth term inside the curly braces, as $\mu=1$) may be obtained for ${}^m{\hskip -1pt}\Pi$ from the second equation (\ref{eq:Pi-eq+wmw1*}). However, the last term inside the curly braces in (\ref{eq:Pi-eq*0+*}) may also be omitted. For this, we note that $\omega=kc$ and observe from (\ref{eq:n_n-ss}) that $\omega^2_{\tt p}$ is expressed in terms of various inverse powers of $r$. We may introduce the static, spherically symmetric plasma potential $V_{\tt p}(r)$, with the terms that decay either as $ r^{-2}$ or faster:
{}
\begin{eqnarray}
\hskip -30pt
V_{\tt p}(r)&=&\frac{\omega_{\tt p}^2(r)}{c^2}+\frac{(\omega_{\tt p}^2)''}{4k^2c^2}=\frac{4\pi e^2}{m_ec^2}
\sum_i \alpha_i \Big(\frac{R_\odot}{r}\Big)^{\beta_i}\Big\{1+
\frac{\beta_i(\beta_i+1)}{4k^2R_\odot^2}
\Big(\frac{R_\odot}{r}\Big)^2\Big\}.
\label{eq:V-sr*}
\end{eqnarray}

The two terms in the curly braces in (\ref{eq:V-sr*}) represent the repulsive potentials due to plasma that, based on the model given by Eq.~(\ref{eq:n_n-ss}), vanish as $r^{-2}$ or faster.  The second plasma term in this expression is dominated by a factor of $(kR_\odot)^{-2}$, which, given the large value of the solar radius, makes its contribution negligible, especially at optical wavelengths ($\lambda\sim1~\mu$m), for which $(kR_\odot)^{-2}\sim 5.23\times 10^{-32}$. Therefore, the term $\propto (\omega_{\tt p}^2)''$ in (\ref{eq:V-sr*}) may be neglected. Although the remaining terms are also small, they may contribute to the phase shifts of the scattered wave and, therefore, they may affect the diffraction of light by the Sun. Thus, they  will be considered. Therefore, the plasma potential, $V_{\tt p}(r)$, in (\ref{eq:Pi-eq*0+*}) has the following from
{}
\begin{eqnarray}
V_{\tt p}(r)&=&
\frac{4\pi e^2}{m_ec^2}
\sum_i \alpha_i \Big(\frac{R_\odot}{r}\Big)^{\beta_i} +{\cal O}\big((kR_\odot)^{-2}\big).
\label{eq:V-sr}
\end{eqnarray}

As a result, and taking into account that $\mu$ is constant, both equations (\ref{eq:Pi-eq+wew1*+}) take an identical form:
{}
\begin{eqnarray}
\Big(\Delta +k^2-V_{\tt p}({ r})\Big)\Pi({\vec r})=0,
\label{eq:Pi-eq*0+*+}
\end{eqnarray}
where the quantity  $\Pi$ represents either the electric Debye potential, ${\,{}^e{\hskip -1pt}\Pi}/{\sqrt{\epsilon}}$, or its magnetic counterpart, ${\,{}^m{\hskip -1pt}\Pi}/{\sqrt{\mu}}$, namely $\Pi({\vec r})=\big({{}^e{\hskip -1pt}\Pi}/{\sqrt{\epsilon}}; {\,{}^m{\hskip -1pt}\Pi}/{\sqrt{\mu}}\big)$, while the plasma potential $V_{\tt p}({ r})$ is given by (\ref{eq:V-sr}).

Equation (\ref{eq:Pi-eq*0+*+}), together with the potential given in Eq.~(\ref{eq:V-sr}), may now be used to determine the solution for the Debye potentials. Together with (\ref{eq:Dr-em0})--(\ref{eq:Bp-em0}), these Debye potentials determine all the components of the EM field.

Note that  (\ref{eq:Pi-eq*0+*+}) resembles the time-independent Schr\"odinger equation describing scattering problems in quantum mechanics \cite{Messiah:1968,Landau-Lifshitz:1989,Burke:2011}. Interestingly, various forms of the power law potential (\ref{eq:V-sr}) appear in many problems of modern atomic physics, related to the scattering of light on a cloud of cold atoms \cite{Friedrich-book-2006,Friedrich-book-2013}. Our method is developed from a generic case of spherically symmetric potentials, many of which are found in the literature describing atomic collisions \cite{Gribakin-Flambaum:1993,Flambaum-etal:1999,Mueller-etal:2011,Friedrich-Trost:2004}. The tools developed in our present paper may also be applicable to these problems in atomic physics.

\section{Solution for the EM field}
\label{sec:sol-EM-Deb}

As we discussed above, to find the solution to the Maxwell equations (\ref{eq:rotE_fl}), we first have to solve (\ref{eq:Pi-eq*0+*+}) for the Debye potential and then use the result in (\ref{eq:Dr-em0})--(\ref{eq:Bp-em0}) to obtain each component of the EM field.
Typically \cite{Born-Wolf:1999}, in spherical polar coordinates, the solution to Eq.~(\ref{eq:Pi-eq*0+*+}) is obtained by separating variables:
\begin{eqnarray}
\Pi({\vec r})=\frac{1}{r}R(r)\Theta(\theta)\Phi(\phi),
\label{eq:Pi*}
\end{eqnarray}
with coefficients that are determined by boundary conditions. Direct substitution into (\ref{eq:Pi-eq*0+*}) reveals that the functions $R, \Theta$ and $\Phi$ must satisfy the following ordinary differential equations:
{}
\begin{eqnarray}
&&\frac{d^2 R}{d r^2}+\Big(k^2 -\frac{\alpha}{r^2}-V_{\tt p}({r})\Big)R=0,
\label{eq:R-bar*}\\
&&\frac{1}{\sin\theta}\frac{d}{d \theta}\Big(\sin\theta \frac{d \Theta}{d \theta}\Big)+\big(\alpha-\frac{\beta}{\sin^2\theta}\big)\Theta=0,
\label{eq:Th*}\\
&&\frac{d^2 \Phi}{d \phi^2}+\beta\Phi=0.
\label{eq:Ph*}
\end{eqnarray}

The solution to (\ref{eq:Ph*}) is given as usual \cite{Born-Wolf:1999}:
{}
\begin{eqnarray}
\Phi_m(\phi)=\Phi_0e^{\pm im(\phi-\phi_0)}  \quad\rightarrow \quad \Phi_m(\phi)=a_m\cos (m\phi) +b_m\sin (m\phi),
\label{eq:Ph_m}
\end{eqnarray}
where $m$ is an integer, and $a_m$ and $b_m$ are integration constants.

Equation (\ref{eq:Th*}) is well known for spherical harmonics. Single-valued solutions to this equation exist when $\alpha=l(l+1)$ with ($l>|m|,$ integer). With this condition, the solution to (\ref{eq:Th*}) becomes
{}
\begin{eqnarray}
\Theta_{lm}(\theta)&=&P^{(m)}_l(\cos\theta).
\label{eq:Th_lm}
\end{eqnarray}

We now focus on the equation for the radial function (\ref{eq:R-bar*}), where, because of (\ref{eq:Th*}), we have $\alpha=\ell(\ell+1)$.  As a result, (\ref{eq:R-bar*}) takes the form
{}
\begin{eqnarray}
\frac{d^2 R}{d r^2}+\Big(k^2-\frac{\ell(\ell+1)}{r^2}-V_{\tt p}(r)\Big)R&=&0.
\label{eq:R-bar-k*}
\end{eqnarray}

To determine the solution to (\ref{eq:R-bar-k*}), we first separate the terms in the plasma potential (\ref{eq:V-sr}) by isolating the $1/r^2$ term from the rest of the terms in the plasma potential (calling it the short-range potential $V_{\tt sr}$) and present (\ref{eq:V-sr})  as
{}
\begin{eqnarray}
\hskip -60pt
V_{\tt p}(r)&=&
\frac{\mu^2}{r^2}+ V_{\tt sr}(r),
\label{eq:V-sr-m20}
\quad {\rm with} \quad
\mu^2=\frac{4\pi e^2R^2_\odot}{m_ec^2} \alpha_2,
\quad
V_{\tt sr}(r)\,=\,\frac{4\pi e^2}{m_ec^2}
\sum_{i>2} \alpha_i \Big(\frac{R_\odot}{r}\Big)^{\beta_i},
\label{eq:V-sr-m2}
\end{eqnarray}
where $\mu^2$ is\footnote{Note the reuse of the symbol $\mu$, do not confuse it with magnetic permeability.} the strength of the $1/r^2$ term in the plasma model at $r=R_\odot$. Using the values from the phenomenological model (\ref{eq:model}),
we can evaluate this term: $\mu^2\simeq 5.89\times 10^{15}$.  The range of $V_{\tt sr}$ is very short; this provides a negligible contribution after $r\simeq 8 R_\odot$. Nevertheless, as it propagates through the solar system, light acquires the largest phase shift as it travels through the range of validity of this potential. Thus, it is important to keep $V_{\tt sr}$  in the model.

The separation of the terms in the plasma potential (\ref{eq:V-sr-m20})
allows us to present the radial equation (\ref{eq:R-bar-k*}) as
{}
\begin{eqnarray}
\frac{d^2 R_L}{d r^2}+\Big(k^2-\frac{L(L+1)}{r^2}- V_{\tt sr}(r)\Big)R_L&=&0,
\label{eq:R-bar-k*2}
\label{eq:R-bar-k-sr}
\end{eqnarray}
where the new index $L$ is determined from
\begin{eqnarray}
\hskip -30pt
L(L+1)=\ell(\ell+1)+\mu^2
\label{eq:L}
\qquad \Rightarrow \qquad
L=\ell+\frac{\mu^2}{\sqrt{(\ell+{\textstyle\frac{1}{2}})^2+\mu^2}+\ell+{\textstyle\frac{1}{2}}}.
\label{eq:L2}
\end{eqnarray}
The solution for $L$  above was obtained under condition that when $\mu\rightarrow0$, the new index $L$ must behave as $L\rightarrow \ell$.
When $\mu/\ell\ll1$, this solution behaves as
{}
\begin{equation}
L\approx \ell+\frac{\mu^2}{2\ell+1} +{\cal O}(\mu^{4}/\ell^3).
\label{eq:L2-apr}
\end{equation}

For a typical region where the plasma potential (\ref{eq:n-eps_n-ism}) is present,  the value of $\ell$ may be estimated using its relation to the classical impact parameter, namely $\ell=kb\geq kR_\odot=4.37\times 10^{15}$. Therefore, the quantity $\mu/\ell\leq 1.75\times 10^{-8}$ is indeed small, justifying the approximation (\ref{eq:L2-apr}).

\subsection{Eikonal solution for Debye potential}
\label{sec:eik}

We address the scattering of high frequency EM waves on the plasma-induced potential  $V_{\tt p}$ that i ed by the heliocentric distance to the heliopause, $R_\star$ from  (\ref{eq:n-eps_n-ism}). In this case and for the case of high energy scattering, we implement the so-called eikonal (or high-energy) approximation \cite{Akhiezer-Pomeranchuk:1950,Glauber-Matthiae:1970,Semon-Taylor:1977,Sharma-etal:1988,Sharma-Somerford:1990,Sharma-Sommerford-book:2006}. In this approximation, the short-range plasma potential contributes a phase shift to the EM wave which can be directly calculated.

\subsubsection{Solution with short-range potential $V_{\tt sr}$  absent.}
\label{sec:short-range=0}

Eq.~(\ref{eq:Pi-eq*0+*+}) can be solved numerically, but only with a great deal of effort, especially at large energies. An exact closed form solution for Eq.~(\ref{eq:Pi-eq*0+*+}) for the general case $V_{\tt sr}\ne 0$ does not exist. However, a number of approximation methods to solve equations of this type were developed for scattering problems in quantum mechanics. At large incident energies, for a wavefront moving in the forward direction, a very useful approximation becomes available. This is the eikonal approximation \cite{Akhiezer-Pomeranchuk:1950,Glauber-Matthiae:1970,Semon-Taylor:1977,Sharma-etal:1988,Sharma-Somerford:1990,Sharma-Sommerford-book:2006}. The eikonal approximation is valid when the following two criteria are satisfied \cite{Sharma-Sommerford-book:2006}:
$kb\gg 1$ and $V_{\tt sr}(r)/k^2\ll 1$. In our case, both of these conditions are fully satisfied, indeed, the first condition yields
$kb=4.37\times 10^{15}\,(\lambda/1\,\mu{\rm m})(b/R_\odot) \gg 1$ and also, taking the short-range plasma potential $V_{\tt sr}$ from (\ref{eq:V-sr-m2}), we evaluate  the second condition as $V_{\tt sr}(r)/k^2\leq V_{\tt sr}(R_\odot)/k^2\approx 4.07\times 10^{-13} \,(\lambda/1\,\mu{\rm m})^2\ll1$.

To develop a solution to (\ref{eq:Pi-eq*0+*+}) using the eikonal approximation, we first note that when $V_{\tt sr}=0$, (\ref{eq:R-bar-k*2})  takes the form
{}
\begin{eqnarray}
\frac{d^2 R_L}{d r^2}+\Big(k^2-\frac{L(L+1)}{r^2}\Big)R_L&=&0.
\label{eq:R-bar-k*20}
\end{eqnarray}
The solution to this equation is well known and is given in terms of Riccati--Bessel functions \cite{Born-Wolf:1999,Turyshev-Toth:2018}:
{}
\begin{eqnarray}
R^{(2)}_L=c_L\psi_L(kr)+d_L\chi_L(kr),
\end{eqnarray}
where the subscript ${}^{(2)}$ simply stands for the solution to (\ref{eq:R-bar-k*20}) that includes the inverse-square term, $1/r^2$.
With the solution for $R^{(2)}_L$ is known, we combine results for $\Phi(\phi)$, $\Theta(\theta)$, given by (\ref{eq:Ph_m}) and (\ref{eq:Th_lm}), to obtain the corresponding Debye potential, $\Pi^{(2)}({\vec r})$, in the form
{}
\begin{eqnarray}
\hskip -25pt
\Pi^{(2)}({\vec r})&=&\frac{1}{r} \sum_{\ell=0}^\infty\sum_{m=-\ell}^\ell
\mu_\ell R^{(2)}_L(r)\big[ P^{(m)}_l(\cos\theta)\big]\big[a_m\cos (m\phi) +b_m\sin (m\phi)\big],
\label{eq:Pi-degn-sol-00}
\end{eqnarray}
where $L=L(\ell)$ is given by (\ref{eq:L2}) and $\mu_\ell, a_m, b_m$ are arbitrary and as yet unknown constants to be determined later.
This solution is well-known and can be studied with available analytical tools (e.g., \cite{Born-Wolf:1999}).

Examining (\ref{eq:Pi-eq*0+*+}), we see that $\Pi^{(2)}({\vec r})$ is a solution to the following wave equation:
{}
\begin{eqnarray}
\Big(\Delta +k^2-\frac{\mu^2}{r^2}\Big)\Pi^{(2)}({\vec r})=0,
\label{eq:Pi-eq*0+*+0}
\end{eqnarray}
which is the equation for the Debye potential $\Pi^{(2)}({\vec r})$ that is as yet unperturbed by the short-range potential, $V_{\tt sr}$.

It is also useful to  explore an approximate solution to (\ref{eq:R-bar-k*20}).
Following \cite{Turyshev-Toth:2017,Herlt-Stephani:1976}, we do that by using the  Wentzel--Kramers--Brillouin (WKB) approximation. In the case when $k$ is rather large (for optical wavelengths $k=2\pi/\lambda=6.28\cdot10^6\,{\rm m}^{-1}$) or when $k\rightarrow\infty$, we established the following asymptotic expression for the radial function   $R_\ell$, valid to order of ${\cal O}\big((kr)^{-5}\big)$:
{}
\begin{eqnarray}
\hskip -72pt
R_\ell(r)&\sim&
\exp\Big[{\frac{\ell(\ell+1)}{4k^2r^2}}+{\frac{[\ell(\ell+1)]^2}{8k^4r^4}}\Big]
\exp\Big[\pm i\Big(kr-\frac{\pi \ell}{2}+\frac{\ell(\ell+1)}{2kr}+\frac{\big[\ell(\ell+1)\big]^2}{24k^3r^3}
\Big)\Big].
\label{eq:R_solWKB+=_bar-imp}
\end{eqnarray}

In \cite{Turyshev-Toth:2017} the asymptotic behavior of the Riccati--Bessel functions was obtained for very larger distances from the turning point for $r\gg r_{\tt t}$; the solution (\ref{eq:R_solWKB+=_bar-imp}) improves them by extending the argument of these functions to shorter distances, closer to the turning point. We obtain similar expressions from the asymptotic expansions of the Riccati--Bessel functions given as finite sums \cite{Korn-Korn:1968,Kerker-book:1969}, to be used in our approach.

\subsubsection{Eikonal wavefunction.}
\label{sec:eik-wfr}

We may now proceed with solving  (\ref{eq:Pi-eq*0+*+}), given the relevant form of $V_{\tt sr}$, (\ref{eq:V-sr-m2}), first representing this equation as
{}
\begin{eqnarray}
\Big(\Delta +k^2-\frac{\mu^2}{r^2}-V_{\tt sr}({ r})\Big)\Pi({\vec r})=0.
\label{eq:Pi-eq*0+*+1*}
\end{eqnarray}

To apply the eikonal approximation to solve this equation, we consider a trial solution of (\ref{eq:Pi-eq*0+*+1*}) in the form
{}
\begin{eqnarray}
\Pi({\vec r})=\Pi^{(2)}({\vec r})\phi(\vec r).
\label{eq:Pi-eq*0+*+1}
\end{eqnarray}
In other words, in the eikonal approximation the Debye potential $\Pi^{(2)}(\vec r)$, becomes ``distorted'' in the presence of the potential $V_{\tt sr}$ (\ref{eq:V-sr-m2}), by $\phi$, a slowly varying function of $r$, such that
{}
\begin{equation}
\left| \nabla^2 \phi \right|\ll k\left|\nabla \phi\right|.
\label{eq:eik2h+}
\end{equation}

When substituted in (\ref{eq:Pi-eq*0+*+1*}), the trial solution (\ref{eq:Pi-eq*0+*+1}) yields
{}
\begin{eqnarray}
&&\Big\{\Delta \Pi^{(2)}({\vec r})+\Big(k^2-\frac{\mu^2}{r^2}\Big)\Pi^{(2)}({\vec r})\Big\}\phi({\vec r})+\nonumber\\[3pt]
&&+2\big({\vec \nabla}\Pi^{(2)}({\vec r})\cdot{\vec \nabla} \phi({\vec r})\big)+\Pi^{(2)}({\vec r})\Delta \phi({\vec r})- V_{\tt sr}({\vec r})\Pi^{(2)}({\vec r})\phi(\vec r)=0.
\label{eq:eik4h0-nab}
\end{eqnarray}

As $\Pi^{(2)}({\vec r})$ is the solution of the homogeneous equation (\ref{eq:Pi-eq*0+*+0}), the first term in (\ref{eq:eik4h0-nab}) is zero. Then, neglecting the third term because of (\ref{eq:eik2h+}), we have
{}
\begin{eqnarray}
\Big({\vec \nabla}\ln \Pi^{(2)}({\vec r})\cdot {\vec \nabla} \ln \phi(\vec r)\Big)= {\textstyle\frac{1}{2}} V_{\tt sr}({\vec r}).
\label{eq:eik4h}
\end{eqnarray}

As we discussed above, the plasma contribution is rather small and it is sufficient  to keep only terms that are first order in $\omega^2_{\tt p}/\omega^2$. Thus, to formally solve (\ref{eq:eik4h}) we may present the solution for $\Pi^{(2)}({\vec r})$ from (\ref{eq:Pi-degn-sol-00}) in series form, in terms of the small parameter $\mu/\ell$, which enters via index $L$ as shown in (\ref{eq:L2}). Then, it is sufficient to take only the zeroth order term (i.e., with $\mu=0$) in $\Pi^{(2)}({\vec r})$. It is easier, however, to obtain such a solution directly from  (\ref{eq:Pi-eq*0+*+0}) by setting $\mu=0$, which yields the well-known free space solution, $e^{i(\vec k\cdot \vec r)}$. As a result, to the  accuracy needed to solve  (\ref{eq:eik4h}), we have $\Pi^{(2)}({\vec r})=e^{i(\vec k\cdot \vec r)} +{\cal O}(\omega^2_{\tt p}/\omega^2)$, which allows us to present (\ref{eq:eik4h}) as
{}
\begin{eqnarray}
i({\vec k} \cdot {\vec \nabla}) \ln \phi= {\textstyle\frac{1}{2}} V_{\tt sr}+{\cal O}(\omega^4_{\tt p}/\omega^4).
\label{eq:eik4h**}
\end{eqnarray}

We may now compute the eikonal phase. For this, we need to introduce a derivative along the propagation path. As we discussed  in \cite{Turyshev-Toth:2017}, we represent the unperturbed trajectory of a light ray as
{}
\begin{eqnarray}
\vec{r}(t)&=&\vec{r}_{0}+\vec{n}c(t-t_0)+{\cal O}(\omega_{\tt p}^2/\omega^2),
\label{eq:x-Newt}
\end{eqnarray}
where $\vec n$ is the unit vector on the incident direction of the light ray's propagation path, such that ${\vec k}=k{\vec n}$, ${\vec n}^2=1$, and $\vec r_0$ represents the starting point. Following \cite{Kopeikin:1997,Kopeikin-book-2011,Turyshev-Toth:2017}, we define ${\vec b}=[[{\vec n}\times{\vec r}_0]\times{\vec n}]$ to be the impact parameter of the unperturbed trajectory of the light ray. The vector ${\vec b}$ is directed from the origin of the coordinate system toward the point of the closest approach of the unperturbed path of light ray to that origin. We will use the $z=z(t)$ coordinate of the ray as the parameter along the path: $z =({\vec n}\cdot {\vec r})=({\vec n}\cdot {\vec r}_{0})+c(t-t_0)$,
which may take both negative and positive signs.
These quantities allow us to rewrite (\ref{eq:x-Newt}) as
{}
\begin{eqnarray}
\hskip -50pt
{\vec r}(z)&=&{\vec b}+{\vec n} z+{\cal O}(\omega_{\tt p}^2/\omega^2),
\quad {\rm with} \quad ||{\vec r}(z)|| \equiv r(z) =\sqrt{b^2+z^2}+{\cal O}(\omega_{\tt p}^2/\omega^2).
\label{eq:b}
\end{eqnarray}

As was shown in \cite{Turyshev-Toth:2017}, the differential operator on the left side of (\ref{eq:eik4h**})  is the derivative along the light ray's propagation path, namely  $({\vec k} \cdot {\vec \nabla})\equiv k({\vec n} \cdot {\vec \nabla})=kd/dz$, where $z$ is a parameter taken along the path, which from (\ref{eq:b}) is given as ${\vec r}=({\vec b},z)$. As a result, for (\ref{eq:eik4h**}) we have
{}
\begin{equation}
\frac{d\ln\phi^\pm}{dz} =\pm\frac{1}{2ik} V_{\tt sr}+{\cal O}(\omega^2_{\tt p}/\omega^2),
\label{eq:eik4h*}
\end{equation}
the solutions of which are
 \begin{eqnarray}
\phi^\pm(\vec b)&=&\exp\Big\{\mp\frac{i}{2k}\int^{z}_{z_0}  V_{\tt sr}({\vec b},z') dz' \Big\}.
\label{eq:eik5h}
\end{eqnarray}
That is, we have the following two particular eikonal solutions of (\ref{eq:Pi-eq*0+*+1*}) for $\Pi(\vec r)$:
\begin{equation}
\Pi(\vec r)=\Pi^{(2)}(\vec r)\exp\Big\{\pm i \xi_b(z) \Big\}+{\cal O}(\omega^4_{\tt p}/\omega^4),
\label{eq:eik6h}
\end{equation}
where we introduced the eikonal phase
\begin{equation}
\xi_b(z) =-\frac{1}{2k}\int^{z}_{z_0}  V_{\tt sr}({\vec b},z') dz'.
\label{eq:eik7h}
\end{equation}
Given $V_{\tt sr}({\vec r})$ from (\ref{eq:V-sr-m2}), we reduced the problem to evaluating a single integral to determine the Debye potentiual $\Pi({\vec r})$ from (\ref{eq:Pi-eq*0+*+1}), which is a great simplification of the problem. Given the fact that ${\vec b}$ is constant and by taking the short-range plasma potential $V_{\tt sr}({\vec r})$ from (\ref{eq:V-sr-m2}), we evaluate (\ref{eq:eik7h}) as
{}
\begin{eqnarray}
\xi_b(r)
&=& -\frac{2\pi e^2R_\odot}{m_ec^2k}
\sum_{i>2}\alpha_i\Big(\frac{R_\odot}{b}\Big)^{\beta_i-1}\Big\{Q_{\beta_i}(z)-Q_{\beta_i}(z_0)\Big\},
\label{eq:delta-D*-av0WKB+1*}
\end{eqnarray}
where we introduced the function $Q_{\beta_i}(z)$, which, with $z=({\vec n}\cdot {\vec r})=\sqrt{r^2-b^2}$, is given as
\begin{eqnarray}
Q_{\beta_i}(z)={}_2F_1\Big[{\textstyle\frac{1}{2}},{\textstyle\frac{1}{2}}\beta_i,{\textstyle\frac{3}{2}},-\frac{z^2}{b^2}\Big]\frac{z}{b},
\label{eq:Q}
\end{eqnarray}
with ${}_2F_1[a,b,c,z]$ being the hypergeometric function \cite{Abramovitz-Stegun:1965}. For $r=b$, the function (\ref{eq:Q}) is well-defined, taking the value of $Q_{\beta_i}(0)=0$, for each $\beta_i$. For $r>b$, for any given value of $\beta_i$, the function $Q_{\beta_i}(z)$ rapidly approaches a limit:
\begin{equation}
\lim\limits_{r\to\infty}Q_{\beta_i}\big(\sqrt{r^2-b^2}\big)
=Q^\star_{\beta_i}\equiv \frac{{\textstyle\frac{1}{2}}\beta_i}{\beta_i-1}B[{\textstyle\frac{1}{2}}\beta_i+{\textstyle\frac{1}{2}},{\textstyle\frac{1}{2}}],
\label{eq:eik1hQ}
\end{equation}
where $B[x,y]$ is Euler's beta function.
For the values of $\beta_i$ used in the model (\ref{eq:model}) for the electron number density in the solar corona, these values are:
\begin{eqnarray}
Q^\star_{2}=\frac{\pi}{2}, \qquad Q^\star_{6}=\frac{3\pi}{16}, \qquad Q^\star_{16}=\frac{429\pi}{4096}.
\label{eq:Q-val}
\end{eqnarray}
Note that the quantities $Q_{\beta_i}$ for $\beta_i>2$ are always small, $0\leq |Q_{\beta_i}|<1$, and as functions of $r$, they reach their asymptotic values quite rapidly after $r\simeq3.2 b$.

Next, we place the source at a large distance from the Sun: $|z_0|\gg R_\star$.
Then, from definition (\ref{eq:Q}) and the asymptotic behavior given by (\ref{eq:eik1hQ}), we have $Q_{\beta_i}(z_0)=-Q^\star_{\beta_i}$. As a result, we express the total eikonal phase shift acquired by the wave along its path through the solar system (\ref{eq:delta-D*-av0WKB+1*}) as
{}
\begin{eqnarray}
\xi^{\rm path}_b(r)
&=& -\frac{2\pi e^2R_\odot}{m_ec^2k}
\sum_{i>2}\alpha_i\Big(\frac{R_\odot}{b}\Big)^{\beta_i-1}\Big\{Q^\star_{\beta_i}+Q_{\beta_i}\big(\sqrt{r^2-b^2}\big)\Big\}.
\label{eq:delta-D*-av0WKB+p}
\end{eqnarray}

Expression (\ref{eq:delta-D*-av0WKB+p}) is the total phase shift induced by the short-range plasma potential along the entire path of the EM wave as it propagates through the solar system. One may see that, as the light propagates from the source to the point of closest approach to the Sun, it acquires the first part of the phase shift, i.e.,  the term proportional to $Q^\star_{\beta_i}$ in (\ref{eq:delta-D*-av0WKB+p}). As it continues to propagate, the second term in (\ref{eq:delta-D*-av0WKB+p}) kicks-in, providing an additional contribution.

Substituting the total eikonal phase shift $\xi_b(r)$ of (\ref{eq:delta-D*-av0WKB+p}) in (\ref{eq:eik6h}) results in the desired solution for the Debye potential $\Pi(\vec r)$. Effectively, this solution demonstrates that the phase of the EM wave is modified by the short-range plasma potential, as expected from the eikonal approximation. Although (\ref{eq:eik6h}) is the solution to (\ref{eq:Pi-eq*0+*+1*}), it still  has arbitrary constants  $\mu_\ell, a_m, b_m$ present in (\ref{eq:Pi-degn-sol-00}), which must be chosen to satisfy the boundary value problem that we set out to solve: Determine the EM field as it propagates through the solar system with the refractive medium given by (\ref{eq:n-eps_n-ism}).

\subsection{Solution for the radial function $R_L(r)$}
\label{sec:eik2}

We now proceed, following \cite{Turyshev-Toth:2017}, to solve (\ref{eq:Pi-eq*0+*+}) with the help of (\ref{eq:Pi*}). A particular solution ${\Pi}$ is obtained by multiplying together the functions $\Phi(\phi)$, $\Theta(\theta)$, given by (\ref{eq:Ph_m}) and (\ref{eq:Th_lm}) and the solution for $R_L$ from (\ref{eq:R-bar-k-sr}), which leads to a general solution to (\ref{eq:Pi-eq*0+*+}). Thus, if $R_L$ is known, we may obtain the Debye potential in the following form
{}
\begin{eqnarray}
\hskip -20pt
\Pi({\vec r})&=&\frac{1}{r} \sum_{\ell=0}^\infty\sum_{m=-\ell}^\ell
\mu_\ell R_L(r)\big[ P^{(m)}_l(\cos\theta)\big]\big[a_m\cos (m\phi) +b_m\sin (m\phi)\big],
\label{eq:Pi-degn-sol-0}
\end{eqnarray}
where $L=L(\ell)$ is given by (\ref{eq:L2}) and $\mu_\ell, a_m, b_m$ are arbitrary and as yet unknown constants.

Thus, our immediate task is to solve (\ref{eq:R-bar-k-sr}).
In the plasma-free case, the entire plasma potential $V_{\tt p}$ is absent, thus $L=\ell$.  The solution in this case is known and, for the case of EM waves diffracted (i.e., obscured) by a large sphere, was given in \cite{Turyshev-Toth:2018}. In this case, in order to determine the coefficients $\mu_\ell$ in (\ref{eq:Pi-degn-sol-0}), we choose $R_\ell(r)$ to be the regular Bessel function $\psi_\ell(kr)$, and require the resulting EM field to match the incident plane EM wave. As a result, in the vacuum, the solutions for the electric and magnetic potentials of the incident wave, ${}^e{\hskip -1pt}\Pi_0$ and ${}^m{\hskip -1pt}\Pi_0$, may be given in terms of a single potential $\Pi_0(r, \theta)$ (see  \cite{Born-Wolf:1999,Turyshev-Toth:2017} for details):
{}
\begin{eqnarray}
\hskip -74pt
 \Bigg(\hskip -4pt
\begin{array}{cc}
{}^e{\hskip -1pt}\Pi_0/\sqrt{\epsilon} \\
{}^m{\hskip -1pt}\Pi_0/\sqrt{\mu}
\end{array}
\hskip -4pt  \Bigg)&=&   \Bigg(\hskip -4pt
\begin{array}{cc}
\cos\phi \\
\sin\phi  \\
  \end{array} \hskip -4pt  \Bigg)
\Pi_0(r, \theta), \hskip 10pt
\Pi_0 (r, \theta)=
\frac{E_0}{k^2}\frac{1}{r}\sum_{\ell=1}^\infty i^{\ell-1}\frac{2\ell+1}{\ell(\ell+1)}\psi_\ell(kr) P^{(1)}_\ell(\cos\theta).
  \label{eq:Pi_ie*+*=}
\end{eqnarray}

Considering the plasma, we notice that,  for large $r$, the potential $ V_{\tt sr}(r)$ in (\ref{eq:R-bar-k-sr}) can be neglected  and this equation reduces to the vacuum discussed in \cite{Turyshev-Toth:2018} with the solution given by  (\ref{eq:Pi_ie*+*=}). The solution of (\ref{eq:R-bar-k-sr}) that is regular at the origin can thus be written asymptotically as a linear combination of the regular and irregular Riccati--Bessel functions $\psi_L(kr)$ and $\chi_L(kr)$, respectively  \cite{Hull-Breit:1959,Friedrich-book-2006,Friedrich-book-2013,Burke-book-2011}, which are solutions of (\ref{eq:R-bar-k-sr}) in the absence of the potential $V_{\tt sr}(r )$. Asymptotically these functions behave as
{}
\begin{eqnarray}
 \psi_L(kr) &\sim& \sin\Big(kr-\frac{\pi L}{2}\Big),
 \label{eq:F-beh}
 \qquad
 \chi_L(kr) \sim \cos\Big(kr-\frac{\pi L}{2}\Big).
\label{eq:G-beh}
\end{eqnarray}
Hence, we look for a solution satisfying the boundary conditions
{}
\begin{eqnarray}
R_L (r) &
\stackrel[r\rightarrow 0]{}{\sim}
& nr^{\ell+1},
\label{eq:R-anz0}\\
R_L (r) &
\stackrel[r\rightarrow 0]{}{\sim}
&  \psi_L(kr) + \chi_L(kr)\,\tan \delta_\ell
\sim\sin\Big(kr-\frac{\pi L}{2}+\delta_\ell\Big),
\label{eq:R-anz}
\end{eqnarray}
where $n$ is a normalization factor. The quantity $\delta_\ell$ introduced in these equations is the phase shift due to the presence of the short-range potential $V_{\tt sr}(r)$. We note that $\delta_\ell$ vanishes when the short-range potential is absent and, thus, it contains all the information necessary to describe the scattering by $V_{\tt sr}(r)$. We generalize expressions (\ref{eq:R-anz0})--(\ref{eq:R-anz}) in the case when the plasma potential has an additional $1/r^2$ term, leading to the substitution $\ell\rightarrow L$.

We can satisfy the conditions (\ref{eq:R-anz0})--(\ref{eq:R-anz}) by choosing the  function $R_L(r)$  in a form of a linear combination of the two solutions (\ref{eq:eik6h}). We rely on solutions in the form of incident and outgoing waves \cite{Thomson-Nunes-book:2009}, given by the functions $\zeta^{(-)}_L(kr)$ and  $\zeta^{(+)}_L(kr)$, correspondingly, and we can show explicit dependence on the eikonal phase shift, $\xi_b(r)$:
{}
\begin{eqnarray}
R_L (r) = \frac{1}{2i}\Big(\zeta^{(+)}_L(kr)e^{i \xi_b(r)} -\zeta^{(-)}_L(kr)e^{-i \xi_b(r)}\Big),
\label{eq:R-L}
\end{eqnarray}
where $\zeta_L^{(\pm)}(kr)=\chi_L(kr)\pm i\psi_L(kr)$ (for discussion, see Appendix~A of \cite{Turyshev-Toth:2018}) with their asymptotic behavior given by
{}
\begin{equation}
\zeta^{(\pm)}_L(kr)
\stackrel[r\rightarrow 0]{}{\sim}
e^{\pm i\big(kr-\frac{1}{2}\pi L\big)}.
\label{eq:eik1h}
\end{equation}
In addition, $\xi_b(r)$ is the eikonal phase shift that is accumulated by the EM wave starting from the point of closest approach, $r=b$. The expression for the quantity is obtained directly from (\ref{eq:eik7h}) by setting $z_0=0$ (or, equivalently, from (\ref{eq:delta-D*-av0WKB+p}) by dropping the $Q^\star_{\beta_i}$-term),   which results in
{}
\begin{eqnarray}
\xi_b(r)
&=& -\frac{2\pi e^2R_\odot}{m_ec^2k}
\sum_{i>2}\alpha_i\Big(\frac{R_\odot}{b}\Big)^{\beta_i-1}Q_{\beta_i}\big(\sqrt{r^2-b^2}\big).
\label{eq:delta-D*-av0WKB+}
\end{eqnarray}

Clearly, the solution for $R_L$
from (\ref{eq:R-L})  satisfies (\ref{eq:R-bar-k-sr}) with the condition (\ref{eq:eik2h+}). It also satisfies the conditions (\ref{eq:R-anz0})--(\ref{eq:R-anz}). Indeed, as the plasma potential exists only for $R_\odot \leq r\leq R_\star$ (which is evident from (\ref{eq:eps}) and (\ref{eq:n-eps_n-ism})), the eikonal phase $\xi_b$ is zero for $r<R_\odot$. Therefore, as $r\rightarrow 0$, the index $L\rightarrow \ell$ and the radial function  (\ref{eq:R-L}) becomes $R_L (r) \rightarrow \psi_\ell(kr)$. However, we know that $\psi_\ell(kr)$ obeys the condition (\ref{eq:R-anz0}). Next, we consider another limit, when $r\rightarrow \infty$. Using the asymptotic behavior of $\zeta^{(\pm)}_L$ from (\ref{eq:eik1h}), we see that, as $r\rightarrow \infty$, the radial function obeys the asymptotic condition (\ref{eq:R-anz}), taking the form where the phase shift $\delta_\ell$ is given by the eikonal phase $\xi_b$ introduced by (\ref{eq:eik7h}).  As  a result, we established that the radial function (\ref{eq:R-L}) represents a desirable solution to (\ref{eq:R-bar-k-sr}) inside the termination shock, $0\leq r\leq R_\star$.

We may further simplify the result (\ref{eq:R-L}), putting it in the following equivalent form:
{}
\begin{eqnarray}
R_L (r) &=&  \cos\xi_b(r) \, \psi_L(kr)+ \sin\xi_b(r)\,\chi_L(kr),
\label{eq:R-L3}
\end{eqnarray}
which explicitly shows the phase shift induced by the short-range plasma potential.

To match the potentials (\ref{eq:Pi_ie*+*=}) with those of the incident and scattered waves, the latter must be expressed in a similar series form, but with arbitrary coefficients. Only the function $\psi_L(kr)$ may be used in the expression for the potential inside the sphere since $\chi_L(kr)$ diverges at the origin. On the other hand, the scattered wave must vanish at infinity, and the functions $\zeta^{(+)}_L(kr)$ will impart precisely this property. These functions are suitable as representations of a scattered wave. For large values of the argument $(kr)$, the result behaves as  $e^{ikr}$ and the Debye potential will be $\Pi\propto e^{ikr}/r$ for large $r$. Thus, at large distances from the sphere (i.e., beyond the termination shock boundary) the scattered wave is spherical, with its center at the origin $r=0$. Accordingly, it will be used in the expression for the scattered wave, i.e., in the trial solution for the Debye potentials of the scattered wave for $r>R_\star$.

Collecting results for the functions $\Phi(\phi)$, $\Theta(\theta)$ from
 (\ref{eq:Ph_m}), (\ref{eq:Th_lm}),  correspondingly, and $R_L(r)=\zeta^{(+)}_L(kr)e^{i\xi_b(r)}$ from (\ref{eq:eik6h}), we have the Debye potential for the scattered wave:
{}
\begin{eqnarray}
\hskip -50 pt
\Pi_{\tt s}&=&\frac{1}{r} \sum_{\ell=0}^\infty\sum_{m=-\ell}^\ell
a_\ell \zeta^{(+)}_L(kr)e^{i\xi_b(r)}\big[ P^{(m)}_\ell(\cos\theta)\big]\big[a'_m\cos (m\phi) +b'_m\sin (m\phi)\big],
~~~~~
\label{eq:Pi-degn-sol-s}
\end{eqnarray}
where $a_\ell, a'_m, b'_m$ are arbitrary and yet unknown constants and the relation between $L$ and $\ell$ is given by (\ref{eq:L2}).

Representing the potential inside the termination shock via $\psi_L(kr)$ is appropriate. Hence the trial solution to (\ref{eq:Pi-eq*0+*+}) for the electric and magnetic Debye potentials inside the termination shock ($r \leq R_\star$) relies on the radial function $R_L(r)$ given by (\ref{eq:R-L3}) and has the form
{}
\begin{eqnarray}
\Pi_{\tt in}&=&\frac{1}{r} \sum_{\ell=0}^\infty\sum_{m=-\ell}^\ell
b_\ell \Big\{ \cos\xi_b(r)\psi_L(kr) + \sin\xi_b(r)\,\chi_L(kr)\Big\}\times \nonumber\\
&&\hskip 60pt \times
\big[ P^{(m)}_\ell(\cos\theta)\big]\big[a_m\cos (m\phi) +b_m\sin (m\phi)\big],
~~~~~
\label{eq:Pi-degn-sol-in}
\end{eqnarray}
where $b_\ell, a_m, b_m$ are arbitrary and yet unknown constants.
We recall that $\Pi$ relates to the electric, ${\,{}^e{\hskip -1pt}\Pi}/{\sqrt{\epsilon}}$, or magnetic, ${\,{}^m{\hskip -1pt}\Pi}/{\sqrt{\mu}}$, Debye potentials by $\Pi({\vec r})=\big({{}^e{\hskip -1pt}\Pi}/{\sqrt{\epsilon}}; {\,{}^m{\hskip -1pt}\Pi}/{\sqrt{\mu}}\big)$, the expression that was introduced after (\ref{eq:Pi-eq*0+*+}).

Finally, in order for the components ${\hat E}_\theta, {\hat E}_\phi$ and ${\hat H}_\theta, {\hat H}_\phi$ to be continuous over the spherical surface $r=R_\star$ of the termination shock, the boundary conditions \cite{Born-Wolf:1999}, set at the termination shock boundary, $r=R_\star$, for the electron plasma distribution (\ref{eq:eps}) and (\ref{eq:n-eps_n-ism}) with $n_0=0$ and thus with $\epsilon(R_\star)=\mu(R_\star)=1$ have the form \cite{Born-Wolf:1999,Turyshev-Toth:2017}:
{}
\begin{eqnarray}
\frac{\partial }{\partial r}\Big[r\,{}^e{\hskip -1pt}\Pi_0+r\,{}^e{\hskip -1pt}\Pi_{\tt s}\Big]_{r=R_\star}&=&\frac{\partial }{\partial r}\Big[{r\,{}^e{\hskip -1pt}\Pi_{\tt in}}\Big]_{r=R_\star},
\label{eq:bound_cond-expand1+}\\[3pt]
\frac{\partial }{\partial r}\Big[r\,{}^m{\hskip -1pt}\Pi_0+r\,{}^m{\hskip -1pt}\Pi_{\tt s}\Big]_{r=R_\star}&=&\frac{\partial }{\partial r}\Big[{r\,{}^m{\hskip -1pt}\Pi_{\tt in}}\Big]_{r=R_\star},
\label{eq:bound_cond-expand2+}\\[3pt]
\hskip 20pt
\Big[r\,{}^e{\hskip -1pt}\Pi_0+r\,{}^e{\hskip -1pt}\Pi_{\tt s}\Big]_{r=R_\star}&=&\Big[{r\,{}^e{\hskip -1pt}\Pi_{\tt in}}\Big]_{r=R_\star},
\label{eq:bound_cond-expand3+}\\[3pt]
\hskip 15pt
\Big[ r\,{}^m{\hskip -1pt}\Pi_0+r\,{}^m{\hskip -1pt}\Pi_{\tt s}\Big]_{r=R_\star}&=&\Big[ {r\,{}^m{\hskip -1pt}\Pi_{\tt in}}\Big]_{r=R_\star}.
\label{eq:bound_cond-expand4+}
\end{eqnarray}

From the geometric symmetry
of the problem \cite{Born-Wolf:1999} and by applying the boundary conditions (\ref{eq:bound_cond-expand1+})--(\ref{eq:bound_cond-expand4+}), we
set the constants $a_m$ and $b_m$ for the electric Debye potentials as $a_1=1, b_1=0, a_m=b_m=0~{\rm for}~m\geq2$; and for the magnetic Debye potentials as $a_1=0, b_1=1, a_m=b_m=0~{\rm for}~m\geq2$, with identical values for $a'_m$ and $b'_m$.

As a result, the solutions for the electric and magnetic potentials of the scattered wave, ${}^e{\hskip -1pt}\Pi_{\tt s}$ and ${}^m{\hskip -1pt}\Pi_{\tt s}$, may be given in terms of a single potential $\Pi_{\tt s}(r, \theta)$ (see  \cite{Turyshev-Toth:2017} for details), which is
{}
\begin{eqnarray}
\hskip -70pt
  \Bigg(\hskip -4pt \begin{array}{cc}
{}^e{\hskip -1pt}\Pi_{\tt s} \\
{}^m{\hskip -1pt}\Pi_{\tt s}
  \end{array} \hskip -4pt \Bigg) &=&
  \Bigg(\hskip -4pt \begin{array}{cc}
\cos\phi\\
\sin\phi
  \end{array} \hskip -4pt \Bigg)  \,\Pi_{\tt s}(r, \theta),
  \quad {\rm where}  \quad
 \Pi_{\tt s}(r, \theta)=\frac{1}{r} \sum_{\ell=1}^\infty
a_\ell \zeta^{(+)}_L(kr)e^{i\xi_b(r)} P^{(1)}_\ell(\cos\theta). ~~~~~
  \label{eq:Pi_s+}
\end{eqnarray}

In a relevant scattering scenario, the EM wave and the Sun are well separated initially, so the Debye potential for the incident wave can be expected to have the same form as in the pure plasma-free case that is given by (\ref{eq:Pi_ie*+*=}).
Therefore, the Debye potential for the inner region has the form:
\begin{eqnarray}
\Bigg(\hskip -4pt  \begin{array}{cc}
{}^e{\hskip -1pt}\Pi_{\tt in}/\sqrt{\epsilon} \\
{}^m{\hskip -1pt}\Pi_{\tt in}/\sqrt{\mu} \\
  \end{array} \hskip -4pt \Bigg) =
\Bigg(\hskip -4pt  \begin{array}{cc}
\cos\phi\\
\sin\phi
  \end{array}\hskip -4pt \Bigg) \,\Pi_{\tt in}(r, \theta),
  \label{eq:Pi_in+}
\end{eqnarray}
with the potential $\Pi_{\tt in}$ given as
{}
\begin{eqnarray}
\Pi_{\tt in}(r, \theta)&=&\frac{1}{r} \sum_{\ell=1}^\infty
b_\ell \Big\{ \cos\xi_b(r)\psi_L(kr) + \sin\xi_b(r)\,\chi_L(kr)\Big\}P^{(1)}_\ell(\cos\theta).
\label{eq:Pi-in}
\end{eqnarray}

We thus expressed all the potentials in the series form given in (\ref{eq:Pi-degn-sol-0}), and any unknown constants can now be determined easily. If we substitute the expressions (\ref{eq:Pi_ie*+*=}), (\ref{eq:Pi_s+}) and (\ref{eq:Pi_in+})--(\ref{eq:Pi-in}) into the boundary conditions (\ref{eq:bound_cond-expand1+})--(\ref{eq:bound_cond-expand4+}), we obtain the following linear relationships between the coefficients $a_\ell$ and $b_\ell$ from (\ref{eq:Pi_s+}) and (\ref{eq:Pi-in}), correspondingly:
{}
\begin{eqnarray}
\Big[\frac{E_0}{k^2}i^{\ell-1}\frac{2\ell+1}{\ell(\ell+1)}
\psi'_\ell(kr)+a_\ell \Big(\zeta^{(+)}_L(kr)e^{i\xi_b(r)}\Big)'\Big]_{r=R_\star}&=&b_\ell R'_L(r)\Big|_{r=R_\star},
\label{eq:bound-cond*1}\\
\Big[\frac{E_0}{k^2}i^{\ell-1}\frac{2\ell+1}{\ell(\ell+1)}
\psi_\ell(kr)+a_\ell \zeta^{(+)}_L(kr)e^{i\xi_b(r)}\Big]_{r=R_\star}&=&b_\ell R_L(r)\Big|_{r=R_\star},
\label{eq:bound-cond*2}
\end{eqnarray}
where $R_L(r)$ is from (\ref{eq:R-L3}) and $'=d/dr$.  From the definition of the eikonal phase (\ref{eq:eik7h}) we see that
\begin{equation}
\xi'_{b}(r)\big|_{r=R_\star} \approx-\frac{1}{2k} V_{\tt sr}(r)\Big|_{r=R_\star}.
\label{eq:eik7h+}
\end{equation}
For the electron plasma distribution  (\ref{eq:eps}) and (\ref{eq:n-eps_n-ism}), the value $\xi'_{b}(R_\star)$ is extremely small and may be neglected.  Rescaling, for convenience, $a_\ell$ and $b_\ell$  by introducing $\alpha_\ell$ and $\beta_\ell$ as
{}
\begin{eqnarray}
a_\ell=\frac{E_0}{k^2}i^{\ell-1}\frac{2\ell+1}{\ell(\ell+1)}\alpha_\ell \qquad {\rm and}\qquad b_\ell=\frac{E_0}{k^2}i^{\ell-1}\frac{2\ell+1}{\ell(\ell+1)}\beta_\ell,
\label{eq:a-b}
\end{eqnarray}
from (\ref{eq:bound-cond*1})--(\ref{eq:bound-cond*2}), we have:
{}
\begin{eqnarray}
\psi'_\ell(kR_\star)+\alpha_\ell {\zeta^{(+)}_L}'(kR_\star)e^{i\xi_b(R_\star)}&=&\beta_\ell R'_L(R_\star),
\label{eq:bound-cond*1+}\nonumber\\[3pt]
\psi_\ell(kR_\star)+\alpha_\ell \zeta^{(+)}_L(kR_\star)e^{i\xi_b(R_\star)}&=&\beta_\ell R_L(R_\star).
\label{eq:bound-cond*2+}
\end{eqnarray}
Equations (\ref{eq:bound-cond*1+}) may now be solved to determine the two sets of coefficients $\alpha_\ell$ and $\beta_\ell$:
{}
\begin{eqnarray}
\alpha_\ell&=&e^{-i\xi_b(R_\star)}\frac{\psi_\ell(kR_\star)R'_L(R_\star)-\psi'_\ell(kR_\star)R_L(R_\star)}{R_L(R_\star){\zeta_L^{(+)}}'(kR_\star)-R'_L(R_\star)\zeta_L^{(+)}(kR_\star)},
\label{eq:a_l*}\\[3pt]
\beta_\ell&=&\frac{\psi_l(kR_\star){\zeta_L^{(+)}}'(kR_\star)-\psi'_\ell(kR_\star)\zeta^{(+)}_L(kR_\star)}{R_L(R_\star){\zeta_L^{(+)}}'(kR_\star)-R'_L(R_\star)\zeta_L^{(+)}(kR_\star)}.
\label{eq:b_l*}
\end{eqnarray}
Taking into account the asymptotic behavior of all the functions involved, namely (\ref{eq:eik1h}) for $\zeta^{(+)}_L$ and (\ref{eq:F-beh}) for $\psi_L$ and $\chi_L$, we have the
solution for the coefficients $\alpha_\ell$ and $\beta_\ell$:
 {}
\begin{eqnarray}
\hskip -20pt
\alpha_\ell&=&\sin\delta^*_\ell, \qquad \beta_\ell=e^{i\delta^*_\ell},
\qquad {\rm where}\qquad \delta^*_\ell=-\frac{\pi}{2}(L-\ell)+\xi^\star_b,
\label{eq:a_b_del}
\end{eqnarray}
with $\xi^\star_b=\xi_b(R_\star)$
and $\delta^*_\ell$ being the phase shift induced by the plasma to the phase of the EM wave propagating through the solar system as measured at the termination shock, $\delta^*_\ell=\delta_\ell(z^\star)$.

As expected, when the plasma is absent, $L=\ell$ and $\xi_b=0$, the total plasma phase shift vanishes, resulting in $\delta_\ell=0$. However, in the case of scattering by the plasma, $\xi^*_b=\xi_b(R_\star)\not=0$ and $\delta_\ell$ is important.
For large heliocentric distances along the incident direction, for which $r\gg b$, and certainly for the region outside the termination shock, $r>R_\star$, the  eikonal phase shift $\xi_b^\star=\xi_b(R_\star)$, given by  (\ref{eq:delta-D*-av0WKB+}), together with (\ref{eq:eik1hQ}), will be
{}
\begin{eqnarray}
\xi_b^\star\approx -\frac{2\pi e^2R_\odot}{m_ec^2k}
\sum_{i>2}\alpha_iQ_{\beta_i}^\star\Big(\frac{R_\odot}{b}\Big)^{\beta_i-1},
\label{eq:delta-D*-av0WKB}
\end{eqnarray}
which, for any given $b$, is a constant value. In the case when $\mu/\ell\ll1$ and (\ref{eq:L2-apr}) is valid, expression (\ref{eq:a_b_del}) for the plasma-induced delay, to 
${\cal O}(\mu^4/\ell^3)$, takes the form
 {}
\begin{eqnarray}
\delta_\ell^\star=-\frac{\pi}{2}\frac{\mu^2}{2\ell}+\xi_b^\star.
\label{eq:a_b_del-r_mu0}
\end{eqnarray}

We can evaluate the contribution of the plasma to the phase of the EM wave as the wave traverses the solar system.  In the case of the electron number density model (\ref{eq:model}) and from (\ref{eq:a_b_del}), the plasma phase shift $\delta^*_\ell$ in (\ref{eq:a_b_del-r_mu0}) is given as
{}
\begin{eqnarray}
\delta^*_{\ell}&=&-\eta_2\frac{R_\odot}{b}-\eta_6Q^\star_{6}\Big(\frac{R_\odot}{b}\Big)^5-\eta_{16}Q^\star_{16}\Big(\frac{R_\odot}{b}\Big)^{15}+...,
\label{eq:s-d1}
\end{eqnarray}
with $\eta_2, \eta_6$ and $\eta_{16}$ having the form
{}
\begin{eqnarray}
\eta_2&=&\frac{\pi}{2}\frac{2\pi e^2R_\odot}{m_ec^2k}
{\alpha_2},\qquad
\eta_6=\frac{2\pi e^2R_\odot}{m_ec^2k}{\alpha_6},\qquad
\eta_{16}=\frac{2\pi e^2R_\odot}{m_ec^2k}{\alpha_{16}},
\label{eq:delta-etas}
\end{eqnarray}
where, to derive the expression for $\eta_2$, we used $\mu^2$ from (\ref{eq:V-sr-m2}) and  approximated (\ref{eq:a_b_del}) for the case of $\mu/\ell \ll1$ by using (\ref{eq:L2-apr}) with $Q^\star_{\beta_i}$ in the incident direction is given by (\ref{eq:Q-val}).  Note that this approach results in the additional factor of $\pi/2$ (which came from the first term in (\ref{eq:a_b_del})) that is characteristic to the eikonal approximation  (see discussion in \cite{Friedrich-book-2006,Friedrich-book-2013}). To derive  $\eta_6$ and $\eta_{16}$, we used (\ref{eq:delta-D*-av0WKB}). The empirical model for the free electron number density in the solar corona (\ref{eq:model}) results in the following value for the constants $\eta_2, \eta_6$ and $\eta_{16}$  in (\ref{eq:delta-etas}):
{}
\begin{eqnarray}
\hskip -30pt
\eta_2&=&1.06 \,\Big(\frac{\lambda}{1~\mu{\rm m}}\Big),\,\qquad
\eta_6=303.87 \,\Big(\frac{\lambda}{1~\mu{\rm m}}\Big),\qquad
\eta_{16}=586.17 \,\Big(\frac{\lambda}{1~\mu{\rm m}}\Big).
\label{eq:delta-etas-m}
\end{eqnarray}
Starting from $b\simeq3.65 R_\odot$, the contribution from the $\eta_2$ term becomes the largest among the terms in (\ref{eq:s-d1}) and rapidly becomes dominant over the remaining terms for larger impact parameters.

With this analysis and by using the value for $a_\ell$ from (\ref{eq:a-b}), together with $\alpha_\ell$ and $\delta_\ell^\star$ from (\ref{eq:a_b_del}), we determine that the solution for the scattered potential (\ref{eq:Pi_s+}), for $r>R_\star$, takes the form
{}
\begin{eqnarray}
\hskip -30pt
\Pi_{\tt s}(r, \theta)&=& -\frac{E_0}{2k^2} \frac{1}{r}\sum_{\ell=1}^\infty
i^{\ell}\frac{2\ell+1}{\ell(\ell+1)} \zeta^{(+)}_L(kr)e^{i\frac{\pi}{2}(L-\ell)} \Big(e^{2i\delta^*_\ell}-1\Big)P^{(1)}_\ell(\cos\theta).
\label{eq:Pi-s_a**}
\end{eqnarray}

We realize that the total EM field in this region is given as the sum of the incident and scattered waves, $\Pi=\Pi_0+\Pi_{\tt s}$, with potentials given by (\ref{eq:Pi_ie*+*=}) and (\ref{eq:Pi-s_a**}), correspondingly. Also, using the asymptotic behavior of $\zeta^{(+)}_L$ from (\ref{eq:eik1h}) and with the help of (\ref{eq:a_b_del}), we notice that at large distances from the Sun we can write $\zeta^{(+)}_L(kr)e^{i\xi_b}  \approx \zeta^{(+)}_\ell (kr)e^{\i\delta^*_\ell}$. As a result, we present the Debye potential  in the region $r>R_\star$ in the following form:
{}
\begin{eqnarray}
\hskip -75pt
\Pi(r, \theta)&=& \Pi_0+\Pi_{\tt s}=\frac{E_0}{k^2}\frac{1}{r}\sum_{\ell=1}^\infty i^{\ell-1}\frac{2\ell+1}{\ell(\ell+1)}
\Big\{\psi_\ell(kr) + \frac{1}{2i}\big(e^{2i\delta^*_\ell}-1\big) \zeta^{(+)}_\ell(kr)\Big\}P^{(1)}_\ell(\cos\theta).
\label{eq:Pi-s_a1*}
\end{eqnarray}

Now we may consider the solution for the Debye potential inside the termination shock, for $R_\odot \leq r \leq R_\star$. Using the value for $b_\ell$ from (\ref{eq:a-b}), together with $\beta_\ell$ from (\ref{eq:a_b_del}), we determine the solution for the inner Debye potential (\ref{eq:a_b_del}):
{}
\begin{eqnarray}
\hskip -72pt
\Pi_{\tt in}(r, \theta)&=&\frac{E_0}{k^2} \frac{1}{r} \sum_{\ell=1}^\infty
i^{\ell-1}\frac{2\ell+1}{\ell(\ell+1)}e^{i\delta^*_\ell} \Big\{ \cos\xi_b(r)\psi_L(kr) + \sin\xi_b(r)\,\chi_L(kr)\Big\}P^{(1)}_\ell(\cos\theta).
\label{eq:Pi-in+}
\end{eqnarray}

As the solar plasma effect is rather weak, we may use the asymptotic expressions for $\psi_L, \chi_L$ and $\zeta^{(\pm)}_L$ for $r\geq R_\odot$.  Therefore, the radial function  $R_L (r) $ from (\ref{eq:R-L}) (or, equivalently, from (\ref{eq:R-L3})), in the region of heliocentric distances $R_\odot\leq r\leq R_\star$, may be given as
{}
\begin{eqnarray}
R_L (r) &=& \frac{1}{2i}\Big(\zeta^{(+)}_L(kr)e^{i \xi_b (r)} -\zeta^{(-)}_L(kr)e^{-i \xi_b(r)}\Big)\simeq\nonumber\\
&\simeq&
e^{-i \delta_\ell(r)}\Big\{\psi_\ell(kr) + \frac{1}{2i}\big(e^{2i \delta_\ell(r)}-1\big)\zeta^{(+)}_\ell(kr)\Big\},
\label{eq:R-L3+*}
\end{eqnarray}
where $\delta_\ell(r)$ has the form given by (\ref{eq:a_b_del}) where  the eikonal phase $\xi_b$ is given in its original form (\ref{eq:delta-D*-av0WKB+}), namely
{}
\begin{eqnarray}
\delta_\ell(r)=-\frac{\pi}{2}(L-\ell)+\xi_b(r).
\label{eq:a_b_del-r}
\end{eqnarray}
Similarly to (\ref{eq:a_b_del-r_mu0}), in the case when $\mu/\ell\ll1$ and (\ref{eq:L2-apr}) is valid, expression (\ref{eq:a_b_del-r}) takes the form $\delta_\ell(r)=-\frac{\pi}{2}\frac{\mu^2}{2\ell}+\xi_b(r)+
{\cal O}(\mu^4/\ell^3)$.
As  a result, outside the Sun, we may present (\ref{eq:Pi-in+}) in the following equivalent form:
{}
\begin{eqnarray}
\hskip -75pt
\Pi_{\tt in}(r, \theta)=\frac{E_0}{k^2} \frac{1}{r} \sum_{\ell=1}^\infty
i^{\ell-1}\frac{2\ell+1}{\ell(\ell+1)}e^{i\big(\delta^*_\ell -\delta_\ell(r)\big)} \Big\{\psi_\ell(kr)+ \frac{1}{2i}\big(e^{2i\delta_\ell(r)}-1\big)\zeta^{(+)}_\ell(kr)\Big\}P^{(1)}_\ell(\cos\theta).\nonumber\\
\label{eq:Pi-in+sl}
\end{eqnarray}
Note that this solution is valid, in principle, even inside the opaque Sun. Indeed, because of the plasma model (\ref{eq:eps}) and (\ref{eq:n-eps_n-ism}), the phase shift vanishes, $\delta_\ell=0$, and (\ref{eq:Pi-in+sl}) reduces to the plasma-free solution (\ref{eq:Pi_ie*+*=}).

Each term in (\ref{eq:Pi-in+sl}) has the contribution of the ongoing plasma phase shift given as $\delta^*_\ell -\delta_\ell(r)$, where  $\delta^*_\ell$  and $\delta_\ell(r)$  are given by (\ref{eq:a_b_del}) and (\ref{eq:a_b_del-r}), correspondingly, with eikonal phase shifts for the short-range plasma potential $\xi_b(r)$ and $\xi_b^\star$ given by (\ref{eq:delta-D*-av0WKB+}) and (\ref{eq:delta-D*-av0WKB}), respectively. To evaluate these terms, we derive the differential plasma-induced phase shift for the heliocentric ranges  $R_\odot \leq r\leq R_\star$. Defining $\delta^*_\ell -\delta_\ell(r)=\xi_b^\star-\xi_b(r)\equiv
\delta \xi_b(r)$ from (\ref{eq:delta-D*-av0WKB+}) and (\ref{eq:delta-D*-av0WKB}) we compute
{}
\begin{eqnarray}
\delta \xi_b(r)=
-\frac{2\pi e^2R_\odot}{m_ec^2k}
\sum_{i>2} {\alpha_i}\Big(\frac{R_\odot}{b}\Big)^{\beta_i-1}\Big\{Q_{\beta_i}^\star-Q_{\beta_i}\big(\sqrt{r^2-b^2}\big)\Big\}.
\label{eq:delta-diff}
\end{eqnarray}

Clearly, $\delta \xi_b(r)$ is significant only in the immediate vicinity of the Sun, where $r\simeq R_\odot$, but it falls off rapidly for larger distances.
Using the phenomenological model (\ref{eq:model}), we estimate the magnitude of the differential phase shift (\ref{eq:delta-diff}). For this, with the help of (\ref{eq:delta-etas}) and (\ref{eq:delta-etas-m}), expression (\ref{eq:delta-diff}) takes the from
{}
\begin{eqnarray}
\delta \xi_b(r)&=&
-\Big\{586.17\Big(\frac{R_\odot}{b}\Big)^{15}\Big(Q_{16}^\star-Q_{16}\big(\sqrt{r^2-b^2}\big)\Big)+\nonumber\\
&&
\hskip 20pt
+\,
303.87\Big(\frac{R_\odot}{b}\Big)^{5}\Big(Q_{6}^\star-Q_{6}\big(\sqrt{r^2-b^2}\big)\Big)\Big\}\Big(\frac{\lambda}{1~\mu{\rm m}}\Big).
\label{eq:delta-diff-mod}
\end{eqnarray}

Examining this expression, we see that it reaches its largest value for the smallest impact parameter of $b\simeq R_\odot$. However, even for radio waves passing that close to the Sun, the phase shift (\ref{eq:delta-diff-mod}) results in a practically negligible effect. Evaluating for $\lambda \simeq 1$~cm, the delay introduced by (\ref{eq:delta-diff-mod}) at $r=10R_\odot$ is $\delta d_b= \delta \xi_b(r) (\lambda/2\pi)\simeq 1 \lambda$ and rapidly diminishes as $r$ increases.  In fact, at heliocentric distances beyond  $r\simeq 20R_\odot$, even for such rather long wavelengths, the differential phase shift  introduced  by (\ref{eq:delta-diff-mod}) is totally negligible.
Thus,  we may treat $\delta \xi_b(r)\simeq 0$ in (\ref{eq:delta-diff}). In other words, based on the phenomenological model for solar corona (\ref{eq:model}), for $r\gtrsim 20R_\odot$, for most of the practical applications, we have  $\delta_\ell(r)\simeq \delta^*_\ell$. As a result,  the solution for the Debye potential in the solar system (\ref{eq:Pi-in+sl}) is equivalent to (\ref{eq:Pi-s_a1*}).

We have now identified all the Debye potentials involved in the Mie problem, namely the potential $\Pi_0$ given by (\ref{eq:Pi_ie*+*=}) representing the incident EM field, the potential $\Pi_{\tt s}$ from (\ref{eq:Pi-s_a**}) describing the scattered EM field outside the termination shock, $r>R_\star$, and the potential $\Pi_{\tt in}$ from (\ref{eq:Pi-in+sl}) describing it inside the termination shock, $0<r\leq R_\star$.

\section{EM field in the solar system}
\label{sec:diff-large-sph}

The solution for the Debye potentials for the EM wave in the solar system, given by (\ref{eq:Pi-s_a**}), describes the propagation of light in the extended solar corona. The presence of the Sun itself is not yet captured. For this, we need to set additional boundary conditions that describe the interaction of the Sun with the incident radiation.

Boundary conditions representing the opaque Sun were introduced in \cite{Turyshev-Toth:2017,Herlt-Stephani:1976}. Here we use these conditions again. Specifically, to set the boundary conditions, we rely on the semiclassical analogy between the partial momentum, $\ell$, and the impact parameter, $b$, that is given as $\ell=kb$ \cite{Messiah:1968,Landau-Lifshitz:1989}.
We realize that rays with impact parameter $b\le R_\odot$ are absorbed by the Sun and those with $b>R_\odot$ are transmitted (see discussion of that in \cite{Turyshev-Toth:2018}).  The fully absorbing boundary conditions signify that all the radiation intercepted by the Sun is fully absorbed by it and no reflection or coherent reemission occurs. All intercepted radiation will be transformed into some other forms of energy, notably heat. Thus, we require that no scattered waves exist with impact parameter $b\ll R_\odot$ or, equivalently, for $\ell \leq kR_\odot$. It means that we need to subtract the scattered waves from the incident wave for $\ell \leq kR_\odot$, as was discussed in \cite{Turyshev-Toth:2017}.

\subsection{The Debye potential for EM field in the solar system}
\label{sec:bound-c}

To implement the boundary conditions for the EM wave in the solar system, we rely on the representation of the Riccati--Bessel function $\psi_\ell$ via incident, $\zeta^{(+)}_\ell$, and outgoing, $\zeta^{(-)}_\ell$, waves as $\psi_\ell=(\zeta^{(+)}_\ell-\zeta^{(-)}_\ell)/2i$ (discussed in \cite{Turyshev-Toth:2018} and also after (\ref{eq:R-bar-k*20})) and express the Debye potential (\ref{eq:Pi-s_a1*}) as
{}
\begin{eqnarray}
\hskip -30pt
\Pi(r, \theta)&=&
\frac{E_0}{2ik^2}\frac{1}{r}\sum_{\ell=1}^\infty i^{\ell-1}\frac{2\ell+1}{\ell(\ell+1)}
\Big\{e^{2i\delta^*_\ell}\zeta^{(+)}_\ell(kr)-\zeta^{(-)}_\ell(kr)\Big\}P^{(1)}_\ell(\cos\theta).
\label{eq:Pi-s_a1}
\end{eqnarray}

This form of the combined Debye potential is convenient for implementing the fully absorbing  boundary conditions. Specifically, subtracting from (\ref{eq:Pi-s_a1})  the outgoing wave (i.e., $\propto \zeta^{(+)}_\ell$) for the impact parameters $b\leq R_\odot$ or equivalently for $\ell\in[1,kR_\odot]$, we have
 {}
\begin{eqnarray}
\hskip -30pt
\Pi(r, \theta)&=&
\Pi_0(r,\theta)+ \frac{E_0}{2ik^2}\frac{1}{r}\sum_{\ell=1}^\infty i^{\ell-1}\frac{2\ell+1}{\ell(\ell+1)}
\big(e^{2i\delta^*_\ell}-1\big) \zeta^{(+)}_\ell(kr)P^{(1)}_\ell(\cos\theta) -
\nonumber\\
&&\hskip 36 pt -\,
\frac{E_0}{2ik^2}\frac{1}{r}\sum_{\ell=1}^{kR_\odot} i^{\ell-1}\frac{2\ell+1}{\ell(\ell+1)}e^{2i\delta^*_\ell}\zeta^{(+)}_\ell(kr)P^{(1)}_\ell(\cos\theta).
\label{eq:Pi-s_a2}
\end{eqnarray}

This is our main result, valid for all distances and angles. It is a rather complex expression. It requires the tools of numerical analysis to fully explore the resulting EM field \cite{Kerker-book:1969,vandeHulst-book-1981,Grandy-book-2005}. However, in most practically important applications we need to know the field in the forward direction. Furthermore, our main interest is to study the largest plasma impact on light propagation, which corresponds to the smallest values of the impact parameter. We may simplify the result (\ref{eq:Pi-s_a2}) by taking into account the asymptotic behavior of the function $\zeta^{(+)}_\ell(kr)$, considering the field at large heliocentric distances, such that $kr\gg\ell$, where $\ell$ is the order of the Riccati--Bessel function (see p.~631 of \cite{Morse-Feshbach:1953}).
For  $kr\rightarrow\infty $ and also for $r\gg r_{\tt t}=\sqrt{\ell(\ell+1)}/k$ (see \cite{Turyshev-Toth:2017}), such an expression is given by (\ref{eq:R_solWKB+=_bar-imp}) as
\begin{eqnarray}
\hskip -30pt
\lim_{kr\rightarrow\infty} \zeta^{(+)}_\ell(kr)&\sim&
\exp\Big[i\Big(kr-\frac{\pi\ell}{2}+\frac{\ell(\ell+1)}{2kr}
+\frac{[\ell(\ell+1)]^2}{24k^3r^3}\Big)\Big] +{\cal O}\big((kr)^{-5}\big),
\label{eq:Fass*}
\end{eqnarray}
which includes the contribution from the centrifugal potential in the radial equation (\ref{eq:R-bar-k*}) (see e.g., Appendix A in \cite{Turyshev-Toth:2018} or in \cite{Kerker-book:1969}). As a result, expression (\ref{eq:Fass*}) extends the argument of (\ref{eq:eik1h}) to shorter distances, closer to the turning point of the potential (see discussion in Appendix~F of \cite{Turyshev-Toth:2017}). By including the extended centrifugal term in (\ref{eq:Fass*}) we can now properly describe the contribution of the plasma potential to the bending of the light ray's trajectory.

Thus, we may take the approximate behavior of $\zeta^{(+)}_\ell(kr)$ in (\ref{eq:Fass*}) and use (\ref{eq:Pi-s_a2}) to present the solution for the Debye potential outside the termination shock, $r>R_\star$ as
{}
\begin{eqnarray}
\hskip -30pt
\Pi (r, \theta)&=&\Pi_0 (r, \theta)+
\frac{e^{ikr}}{r}\Big\{\frac{E_0}{2k^2}\sum_{\ell=1}^{kR_\odot} \frac{2\ell+1}{\ell(\ell+1)}
e^{i\big(\frac{\ell(\ell+1)}{2kr}+\frac{[\ell(\ell+1)]^2}{24k^3r^3}\big)}
P^{(1)}_\ell(\cos\theta)-\nonumber\\
&-&
\frac{E_0}{2k^2}\sum_{\ell=kR_\odot}^{\infty} \frac{2\ell+1}{\ell(\ell+1)}e^{i\big(\frac{\ell(\ell+1)}{2kr}+\frac{[\ell(\ell+1)]^2}{24k^3r^3}\big)}
\big(e^{i2\delta^*_\ell}-1\big)P^{(1)}_\ell(\cos\theta) \Big\}=\nonumber\\
&=&
\Pi_0 (r, \theta)+\Pi_{\tt bc} (r, \theta)+\Pi _{\tt p} (r, \theta).
  \label{eq:Pi_g+p0+}
\end{eqnarray}

The first term in (\ref{eq:Pi_g+p0+}), $\Pi_0 (r, \theta)$, is the Debye potential of the pure vacuum case that may be derived from the solution obtained in \cite{Turyshev-Toth:2017}. For practical purposes, one may use an exact expression for $\Pi_0$, which may be given as \cite{Turyshev-Toth:2018}:
\begin{eqnarray}
\Pi_0 (r, \theta)&=&
\frac{E_0}{k^2r \sin\theta}\Big(e^{ikr\cos\theta}-e^{ikr}+{\textstyle\frac{1}{2}}(1-\cos\theta)\big(e^{ikr}-e^{-ikr}\big)\Big),
\label{eq:P0}
\end{eqnarray}
which is the Debye potential of the plasma-free EM wave. This solution is always finite and is valid for any angle $\theta$.

The second term, $\Pi_{\tt bc} (r, \theta)$, is due to the fully absorbing boundary conditions and is responsible for the geometric shadow behind the Sun. The third term, $\Pi _{\tt p} (r, \theta)$,  quantifies the contribution of the solar plasma on the scattering of the EM propagating through the solar system, and evaluated at the distance $r> R_\star$. Because of the plasma model (\ref{eq:eps}) and (\ref{eq:n-eps_n-ism}), the last sum in (\ref{eq:Pi_g+p0+}) formally extends only to $\ell=k R_\star$ corresponding to the impact parameter $b=R_\star$. As expected, for $r>R_\star$, the phase shift $\delta_\ell=0$ and the entire plasma-scattered wave vanishes.

With the solution for the Debye potential given in (\ref{eq:Pi_g+p0+}), and with the help of (\ref{eq:Dr-em0})--(\ref{eq:Bp-em0}) (also see \cite{Turyshev-Toth:2017}), we may now compute the EM field in the various regions involved. Given the value of $(\omega_p/\omega)^2$ ($\sim 10^{-2}$ for radio and $\sim 10^{-11}$ for optical wavelengths), we may neglect the effect of the solar plasma, behaving as $\propto 1/r^2$, on the amplitude of the EM wave. This is especially true at large heliocentric distances where the effect of the plasma on the amplitude of the EM wave is negligible.
The plasma will affect the delay of the EM wave, which is fully accounted for by the solution for the Debye potentials. Thus, we can put $\epsilon =\mu =1$ in (\ref{eq:Dr-em0})--(\ref{eq:Bp-em0}) and use the following expressions to construct the EM field in the static, spherically symmetric geometry (see details in \cite{Turyshev-Toth:2017}):
{}
\begin{eqnarray}
\hskip -30pt
  \Bigg( \hskip -4pt \begin{array}{cc}
{ \hat E}_r \\
{ \hat H}_r
  \end{array} \hskip -4pt \Bigg) &=&
   \left( \begin{array}{cc}
\cos\phi \\
\sin\phi
  \end{array} \right) \,e^{-i\omega t}\alpha(r, \theta), ~~~
    \Bigg( \hskip -4pt \begin{array}{cc}
{ \hat E}_\theta \\
{ \hat H}_\theta
  \end{array} \hskip -4pt \Bigg) =
   \Bigg( \hskip -4pt \begin{array}{cc}
\cos\phi \\
\sin\phi
  \end{array} \hskip -4pt \Bigg) \,e^{-i\omega t}\beta(r, \theta), \nonumber\\
  \hskip -30pt
\Bigg( \hskip -4pt \begin{array}{cc}
{ \hat E}_\phi \\
{ \hat H}_\phi
  \end{array} \hskip -4pt \Bigg) &=&  \Bigg( \hskip -4pt \begin{array}{cc}
-\sin\phi \\
\cos\phi
  \end{array} \hskip -4pt \Bigg) \,e^{-i\omega t}\gamma(r, \theta),
  \label{eq:DB-sol00p*}
\end{eqnarray}
with quantities $\alpha, \beta$ and $\gamma$ are computed from the known Debye potential, $\Pi$, as
{}
\begin{eqnarray}
\hskip -30pt
\alpha(r, \theta)&=&
\frac{\partial^2 }{\partial r^2}
\big({r\,{\hskip -1pt}\Pi}\big)+k^2
\big({r\,{\hskip -1pt}\Pi}\big),
\label{eq:alpha*}
\qquad\qquad
\beta(r, \theta)=\frac{1}{ r}
\frac{\partial^2 \big(r\,{\hskip -1pt}\Pi\big)}{\partial r\partial \theta}+\frac{ik\big(r\,{\hskip -1pt}\Pi\big)}{r\sin\theta},
\label{eq:beta*}\nonumber\\
\hskip -30pt
\gamma(r, \theta)&=&\frac{1}{ r\sin\theta}
\frac{\partial \big(r\,{\hskip -1pt}\Pi\big)}{\partial r}+\frac{ik}{r}
\frac{\partial\big(r\,{\hskip -1pt}\Pi\big)}{\partial \theta}.~~~~~
\label{eq:gamma*}
\end{eqnarray}

We may now use these equations to compute the EM field in the region beyond the termination shock.

\subsection{EM field in the shadow region}

In the shadow behind the Sun (i.e., for impact parameters $b\leq R_\odot$) the EM field is represented by the Debye potential of the shadow, $\Pi_{\tt sh}$, which is given by (\ref{eq:Pi_g+p0+}) as
{}
\begin{eqnarray}
\hskip -20pt
\Pi_{\tt sh} (r, \theta)&=&\Pi_{\tt 0} (r, \theta)+\Pi_{\tt bc} (r, \theta)=\nonumber\\
&=&
\Pi_0 (r, \theta)+\frac{e^{ikr}}{r}\frac{E_0}{2k^2}\sum_{\ell=1}^{kR_\odot} \frac{2\ell+1}{\ell(\ell+1)}
e^{i\big(\frac{\ell(\ell+1)}{2kr}+\frac{[\ell(\ell+1)]^2}{24k^3r^3}\big)}
P^{(1)}_\ell(\cos\theta).
  \label{eq:Pi_sh+}
\end{eqnarray}
As was already discussed in \cite{Turyshev-Toth:2018}, there is no EM field in the geometric shadow behind the Sun and, thus, no light other than the classical Poisson--Arago bright spot \cite{Turyshev-Toth:2018-grav-shadow}.

\subsection{EM field outside the shadow}
\label{sec:EM-field}

In the region outside the solar shadow (i.e., for light rays with impact parameters $b>R_\odot$), the EM field is derived from the Debye potential given by the remaining terms in (\ref{eq:Pi_g+p0+}) as
{}
\begin{eqnarray}
\hskip -60pt
\Pi (r, \theta)&=&\Pi_0 (r, \theta)-
\frac{e^{ikr}}{r}\frac{E_0}{2k^2}\sum_{\ell=kR_\odot}^{\infty} \frac{2\ell+1}{\ell(\ell+1)}
e^{i\big(\frac{\ell(\ell+1)}{2kr}+\frac{[\ell(\ell+1)]^2}{24k^3r^3}\big)}
\Big(e^{i2\delta^*_\ell}-1\Big)P^{(1)}_\ell(\cos\theta)=\nonumber \\
\hskip -60pt
&=&\Pi_0 (r, \theta)+\Pi_{\tt p} (r, \theta).
  \label{eq:Pi_g+p0+in*}
\end{eqnarray}

Expression (\ref{eq:Pi_g+p0+in*}) is our main result for the region outside the termination shock, $r>R_\star$. It contains all the information needed to describe the total EM field originating from a plane wave that passed through the entire region of the extended solar corona, characterized by an electron number density (\ref{eq:n-eps_n-ism}) that diminishes as $r^{-2}$ or faster.

The quantity $\Pi_0 (r, \theta)$ in (\ref{eq:Pi_g+p0+in*}) is the Debye potential of the incident given by (\ref{eq:P0}). Using the exact solution for $\Pi_0 (r, \theta)$ given by (\ref{eq:P0}), and with the help of (\ref{eq:alpha*}) and (\ref{eq:DB-sol00p*}), we compute the EM field produced by this potential  \cite{Turyshev-Toth:2018}:
{}
\begin{eqnarray}
\hskip -60pt
  \Bigg( \hskip -4pt \begin{array}{cc}
{  \hat E}^{\tt (0)}_r \\
{  \hat H}^{\tt (0)}_r
  \end{array} \hskip -4pt \Bigg) &=& E_0\sin\theta
\Bigg( \hskip -4pt \begin{array}{cc}
\cos\phi \\
\sin\phi  \\
  \end{array} \hskip -4pt \Bigg) \,e^{i(kz-\omega t)},
  \label{eq:DB-t-pl=p10}
    \Bigg( \hskip -4pt \begin{array}{cc}
{  \hat E}^{\tt (0)}_\theta \\
{  \hat H}^{\tt (0)}_\theta
  \end{array} \hskip -4pt \Bigg) = E_0\cos\theta
\Bigg( \hskip -4pt \begin{array}{cc}
\cos\phi \\
\sin\phi  \\
  \end{array} \hskip -4pt \Bigg) \,e^{i(kz-\omega t)},
  \nonumber\\
  \hskip -60pt
   \Bigg( \hskip -4pt \begin{array}{cc}
{  \hat E}^{\tt (0)}_\phi \\
{  \hat H}^{\tt (0)}_\phi
  \end{array} \hskip -4pt \Bigg) &=&E_0
  \Bigg( \hskip -4pt \begin{array}{cc}
-\sin\phi \\
\cos\phi  \\
  \end{array} \hskip -4pt \Bigg) \,e^{i(kz-\omega t)}.~~~~
  \label{eq:DB-t-pl=p20}
\end{eqnarray}

The quantity  $\Pi_{\tt p} (r, \theta)$ in (\ref{eq:Pi_g+p0+in*}) is the Debye potential for the plasma-scattered wave that has the form
{}
\begin{eqnarray}
\hskip -20pt
  \Pi_{\tt p} (r, \theta)&=&E_0f_{\tt p}(r,\theta)\frac{e^{ikr}}{r},
  \quad~ {\rm where }\nonumber\\
  \hskip -20pt
   f_{\tt p}(r,\theta)&=&-\frac{1}{2k^2}\sum_{\ell=kR_\odot}^\infty \frac{2\ell+1}{\ell(\ell+1)}
  e^{i\big(\frac{\ell(\ell+1)}{2kr}+\frac{[\ell(\ell+1)]^2}{24k^3r^3}\big)}
  \Big(e^{i2\delta^*_\ell}-1\Big)
P^{(1)}_\ell(\cos\theta).
  \label{eq:Pi_ie*+8p}
\end{eqnarray}

To determine the components of the EM field produced by $\Pi_{\tt p} (r, \theta)$,
we use (\ref{eq:Pi_ie*+8p}) in the expressions (\ref{eq:alpha*}) that yield
{}
\begin{eqnarray}
\hskip -50pt
\alpha(r, \theta)&\hskip -20pt=&\hskip -5pt  -E_0\frac{e^{ikr}}{k^2r^2} \sum_{\ell=kR_\odot}^\infty (\ell+{\textstyle\frac{1}{2}})\Big(1
-\frac{1}{ikr}\Big)
e^{i\big(\frac{\ell(\ell+1)}{2kr}+\frac{[\ell(\ell+1)]^2}{24k^3r^3}\big)}\Big(e^{i2\delta^*_\ell}-1\Big)
P^{(1)}_\ell(\cos\theta),
  \label{eq:alpha*1}\nonumber\\
  ~\\
  \hskip -50pt
\beta(r, \theta)&\hskip -20pt=&E_0\frac{e^{ikr}}{ikr} \sum_{\ell=kR_\odot}^\infty \frac{(\ell+{\textstyle\frac{1}{2}})}{\ell(\ell+1)}
e^{i\big(\frac{\ell(\ell+1)}{2kr}+\frac{[\ell(\ell+1)]^2}{24k^3r^3}\big)}
\Big(e^{i2\delta^*_\ell}-1\Big)\times\nonumber\\
&&\hskip 10pt \times\,\Big\{
\frac{\partial P^{(1)}_\ell(\cos\theta)}{\partial \theta}
\Big(1-\frac{\ell(\ell+1)}{2k^2r^2}-\frac{[\ell(\ell+1)]^2}{8k^4r^4}\Big)+\frac{P^{(1)}_\ell(\cos\theta)}{\sin\theta}\Big\},
  \label{eq:beta**1}\\
  \hskip -50pt
\gamma(r, \theta)&\hskip -20pt=&E_0\frac{e^{ikr}}{ikr} \sum_{\ell=kR_\odot}^\infty \frac{(\ell+{\textstyle\frac{1}{2}})}{\ell(\ell+1)}
e^{i\big(\frac{\ell(\ell+1)}{2kr}+\frac{[\ell(\ell+1)]^2}{24k^3r^3}\big)}
\Big(e^{i2\delta^*_\ell}-1\Big)\times\nonumber\\
&&\hskip 10pt \times\,\Big\{
\frac{\partial P^{(1)}_\ell(\cos\theta)}{\partial \theta}
+\frac{P^{(1)}_\ell(\cos\theta)}{\sin\theta}\Big(1-\frac{\ell(\ell+1)}{2k^2r^2}-\frac{[\ell(\ell+1)]^2}{8k^4r^4}\Big)\Big\}.
  \label{eq:gamma**1}
\end{eqnarray}

To establish the solution for the EM field, we evaluate the expressions (\ref{eq:alpha*1})--(\ref{eq:gamma**1}), substituting the results in (\ref{eq:DB-sol00p*}).

\subsection{Solution for the function $\alpha(r,\theta)$ and the radial components of the EM field}
\label{sec:radial-comp}

We begin with the investigation of $\alpha(r,\theta)$ from (\ref{eq:alpha*1}).
To evaluate this expression, we use the asymptotic representation for $P^{(1)}_l(\cos\theta)$ from \cite{Bateman-Erdelyi:1953,Korn-Korn:1968,Kerker-book:1969}, valid when $\ell\to\infty$:
{}
\begin{eqnarray}
\hskip -65pt
P^{(1)}_\ell(\cos\theta)  =
\frac{-\ell}{\sqrt{2\pi \ell \sin\theta}}\Big(e^{i(\ell+\frac{1}{2})\theta+i\frac{\pi}{4}}+e^{-i(\ell+\frac{1}{2})\theta-i\frac{\pi}{4}}\Big)+{\cal O}(\ell^{-\textstyle\frac{3}{2}}) \quad\textrm{for}\quad 0<\theta<\pi.
\label{eq:P1l<}
\end{eqnarray}

This approximation can be used to transform (\ref{eq:alpha*1}) as
{}
\begin{eqnarray}
\hskip -30pt
\alpha(r, \theta)&=&E_0
\frac{e^{ikr}}{k^2r^2}\sum_{\ell=kR_\odot}^\infty \frac{(\ell+{\textstyle\frac{1}{2}})\sqrt{\ell}}{\sqrt{2\pi \sin\theta}}
\Big(1 -\frac{1}{ikr}\Big)\Big(e^{i2\delta^*_\ell}-1\Big)\,
e^{i\big(\frac{\ell(\ell+1)}{2kr}+\frac{[\ell(\ell+1)]^2}{24k^3r^3}\big)}\times\nonumber\\
&&\hskip 60pt\times
\Big(e^{i(\ell+\frac{1}{2})\theta+i\frac{\pi}{4}}+e^{-i(\ell+\frac{1}{2})\theta-i\frac{\pi}{4}}\Big).~~~
\label{eq:P_sum}
\end{eqnarray}

We recognize that for large $\ell\geq kR^\star_\odot$, we may replace $\ell+1\rightarrow \ell$ and $\ell+\textstyle\frac{1}{2}\rightarrow \ell$.
At this point, we may replace the sum in (\ref{eq:P_sum}) with an integral:
{}
\begin{eqnarray}
\hskip -30pt
\alpha(r, \theta)&=&E_0
\frac{e^{ikr}}{k^2r^2}\int_{\ell=kR_\odot}^\infty \hskip-4pt
\frac{\ell\sqrt{\ell}d\ell}{\sqrt{2\pi \sin\theta}}
\Big(1 -\frac{1}{ikr}\Big)\Big(e^{i2\delta^*_\ell}-1\Big)\,
e^{i\big(\frac{\ell^2}{2kr}+\frac{\ell^4}{24k^3r^3}\big)}\times\nonumber\\
&&\hskip 60pt\times
\Big(e^{i(\ell\theta+\frac{\pi}{4})}+e^{-i(\ell\theta+\frac{\pi}{4})}\Big).~~~
\label{eq:Pi_s_exp1}
\end{eqnarray}
We can drop the $1/(ikr)$ term in this expression, as its magnitude $1/(kr)$ times smaller compared to the leading term. The remaining integral may be evaluated by the method of stationary phase \cite{Turyshev-Toth:2017,Herlt-Stephani:1976}, which applies to integrals of the type
{}
\begin{equation}
I=\int A(\ell)e^{i\varphi(\ell)}d\ell, \qquad
\ell\in\mathbb{R},
\label{eq:stp-1}
\end{equation}
where the amplitude $A(\ell)$ is a slowly varying function of $\ell$, while $\varphi(\ell)$ is a rapidly varying function of $\ell$.
The integral (\ref{eq:stp-1}) may be replaced, to good approximation, with a sum over the points of stationary phase, $\ell_0\in\{\ell_{1,2,..}\}$, for which $d\varphi/d\ell=0$. Defining $\varphi''=d^2\varphi/d\ell^2$, we obtain the integral
{}
\begin{equation}
I\simeq\sum_{\ell_0\in\{\ell_{1,2,..}\}} A(\ell_0)\sqrt{\frac{2\pi}{\varphi''(\ell_0)}}e^{i\big(\varphi(\ell_0)+{\textstyle\frac{\pi}{4}}\big)}.
\label{eq:stp-2}
\end{equation}

Because the scattering term $\big(e^{i2\delta^*_\ell}-1\big)$ in (\ref{eq:Pi_s_exp1}) provides two contributions, each with a different phase, we will treat the integral (\ref{eq:Pi_s_exp1}) as the sum of two integrals: one with and one without the contribution from the plasma phase shift $2\delta^\star_\ell$. To demonstrate our approach, we begin with the plasma-free case.

\subsubsection{Evaluating the plasma-free term.}
\label{sec:no-p-r}

For the term in (\ref{eq:Pi_s_exp1}) without the plasma phase shift, the relevant $\ell$-dependent part of the phase is of the form \cite{Turyshev-Toth:2018}
{}
\begin{eqnarray}
\varphi^{[0]}_{\pm}(\ell)=\pm\Big(\ell\theta+{{\frac{\pi}{4}}}\Big)+\frac{\ell^2}{2kr}+\frac{\ell^4}{24k^3r^3} +{\cal O}\big((kr)^{-5}\big).
\label{eq:S-l}
\end{eqnarray}

The phase is stationary when $d\varphi^{[0]}_{\pm}/d\ell=0$, which implies
{}
\begin{equation}
\hskip 20pt
\mp\theta=\frac{\ell}{kr}\Big(1+\frac{\ell^2}{6k^2r^2}\Big)+{\cal O}\big((kr)^{-5}\big), \qquad {\rm or}\qquad
\ell= \mp kr{\theta}\Big(1-{\textstyle\frac{1}{6}}\theta^2\Big)+{\cal O}(\theta^5),
\label{eq:S-l-pri*}
\end{equation}
therefore, we may write the solution for the points of stationary phase
{}
\begin{equation}
\ell_0 = \mp kr \sin\theta+{\cal O}(\theta^5).
\label{eq:S-l-pri}
\end{equation}
Note that by extending the asymptotic expansion of $\zeta^{(+)}_\ell(kr)$ from (\ref{eq:Fass*}) to ${\cal O}((kr)^{-(2n+1)})$ (i.e., using the WKB approximation as was done in developing (\ref{eq:R_solWKB+=_bar-imp})), the validity of (\ref{eq:S-l-pri}) extends to ${\cal O}(\theta^{2n+1})$.

The solution for $\ell_0$ from  (\ref{eq:S-l-pri}) allows us to compute the phase for the points of stationary phase (\ref{eq:S-l}):
{}
\begin{eqnarray}
\varphi^{[0]}_{\pm}(\ell_0)&=&
\pm\textstyle{\frac{\pi}{4}}+kr\Big(-\textstyle{\frac{1}{2}}\theta^2+\textstyle{\frac{1}{24}}\theta^4\Big)+{\cal O}(\theta^6).
\label{eq:S-l2p}
\end{eqnarray}
To calculate $\varphi''(\ell)$ to ${\cal O}(\theta^6)$ as in (\ref{eq:S-l2p}), we need to include in the phase $\varphi^{[0]}_{\pm}(\ell)$ (\ref{eq:S-l}) another term $\propto \ell^6$, which may be taken  from  (\ref{eq:R_solWKB+=_bar-imp}). This allows us to compute $\varphi''(\ell_0)$:
\begin{eqnarray}
\hskip -30pt
\frac{d^2\varphi^{[0]}_{\pm}}{d\ell^2} &=& \frac{1}{kr}\Big(1+\frac{\ell^2}{2k^2r^2}+\frac{3\ell^4}{8k^4r^4}+{\cal O}\big((kr)^{-6}\big)\Big) \quad  {\rm or,~ for~}\ell=\ell_0, {~\rm we ~ have} \nonumber\\
\hskip -30pt
\varphi''(\ell_0)&\equiv& \frac{d^2\varphi^{[0]}_{\pm}}{d\ell^2} \Big|_{\ell=\ell_0} =
\frac{1}{kr}\Big(1+{\textstyle\frac{1}{2}}\theta^2+{\textstyle\frac{5}{24}}\theta^4+{\cal O}(\theta^6)\Big).~~~~~~~
\label{eq:S-l2}
\end{eqnarray}

The remaining integral is easy to evaluate using the method of stationary phase. Before we do that, we need to bring in the amplitude factor for the asymptotic expansion  $\zeta^{(+)}_\ell(kr)$ given by (\ref{eq:Fass*}). This factor, which we denote by $a(\ell)$, is readily available from (\ref{eq:R_solWKB+=_bar-imp}) in the following form:
{}
\begin{eqnarray}
a(\ell)&=&\exp\Big[{\frac{\ell(\ell+1)}{4k^2r^2}}+{\frac{[\ell(\ell+1)]^2}{8k^4r^4}}\Big]+{\cal O}((kr)^{-6}) \qquad {\rm or,~ for~}\ell=\ell_0, \nonumber\\
 a(\ell_0) &=&1+{\textstyle\frac{1}{4}}\theta^2+{\textstyle\frac{7}{96}}\theta^4+{\cal O}(\theta^6).
\label{eq:sf}
\end{eqnarray}
The fact that we did not use it in (\ref{eq:Fass*}) does not affect results of the calculations above. However, as we shall see below, its presence is needed to offset some of the terms present in the phase of (\ref{eq:S-l}). The significance of this term is in the fact that it cancels
the contribution of the $\theta$-dependence in (\ref{eq:S-l2}). Namely, using the result (\ref{eq:sf}), we derive
{}
\begin{equation}
a(\ell_0)\sqrt{\frac{2\pi}{\varphi''(\ell_0)}}=\sqrt{2\pi kr}+{\cal O}(\theta^6).
\label{eq:sf*}
\end{equation}

Now, using (\ref{eq:sf*}), we have the amplitude of the integrand in (\ref{eq:Pi_s_exp1}), for $\ell \gg1$, taking the form
{}
\begin{eqnarray}
\hskip -60pt
A^{[0]}(\ell_0)a(\ell_0)\sqrt{\frac{2\pi}{\varphi''(\ell_0)}}&=&\frac{\ell_0\sqrt{\ell_0}}{\sqrt{2\pi \sin\theta}}
a(\ell_0)\sqrt{\frac{2\pi}{\varphi''(\ell_0)}}=
(\mp1)^{\textstyle\frac{3}{2}} k^2r^2\sin\theta
\Big(1+{\cal O}(\theta^5)\Big),
\label{eq:S-l3p+*}
\end{eqnarray}
where the superscript ${}^{[0]}$ denotes the term with no plasma contribution. 

As a result, the plasma-free part of the expression for the $\delta\alpha^{[0]}(r,\theta)$ from (\ref{eq:Pi_s_exp1}) takes the form
{}
\begin{eqnarray}
\delta\alpha^{[0]}_\pm(r,\theta)&=&
-E_0\sin\theta e^{ikr\cos\theta}+{\cal O}(\theta^5).
\label{eq:Pi_s_exp4+1*}
\end{eqnarray}
The validity of this result may be extended to ${\cal O}(\theta^{2n+1})$ by extending the asymptotic expansion of $\zeta^{(+)}_\ell(kr)$  (\ref{eq:Fass*}) to ${\cal O}((kr)^{-(2n+1)})$. This can be done by following the approach that led to (\ref{eq:R_solWKB+=_bar-imp}) and repeating the derivations above.

\subsubsection{Evaluating the term with plasma contribution.}
\label{sec:yes-p-r}

We now turn our attention to the term in (\ref{eq:Pi_s_exp1}) that contains a plasma phase shift contribution. The relevant $\ell$-dependent part of the phase is given as
{}
\begin{equation}
\varphi^{[\tt p]}_{\pm}(\ell)=\pm\Big(\ell\theta+{\textstyle{\frac{\pi}{4}}}\Big)+\frac{\ell^2}{2kr}+\frac{\ell^4}{24k^3r^3}+2\delta^\star_\ell,
\label{eq:S-l*p}
\end{equation}
with the plasma contribution clearly shown. From the definition (\ref{eq:delta-D*-av0WKB}) and (\ref{eq:a_b_del-r_mu0}), this plasma phase shift is given as
{}
\begin{eqnarray}
2\delta_b^*
&=&-\frac{4\pi e^2R_\odot}{m_ec^2k}\Big\{
\alpha_2 \frac{\pi}{2}\frac{R_\odot}{b}+
\sum_{i>2} {\alpha_i}Q^\star_{\beta_i}\Big(\frac{R_\odot}{b}\Big)^{\beta_i-1}\Big\}\equiv\nonumber\\
& \equiv& -\frac{2\pi e^2R_\odot}{m_ec^2k}
\sum_{i\geq2}\frac{ \alpha_i\beta_i}{\beta_i-1}B[{\textstyle\frac{1}{2}}\beta_i+{\textstyle\frac{1}{2}},{\textstyle\frac{1}{2}}]\Big(\frac{R_\odot}{b}\Big)^{\beta_i-1},
\label{eq:a_b_del-r+-tot*}
\end{eqnarray}
where we used the representation of $Q^\star_{\beta_i}$ via Euler's beta function, as shown by (\ref{eq:eik1hQ}).

The phase (\ref{eq:S-l*p}) is stationary when $d\varphi^{[\tt p]}_{\pm}/d\ell=0$, which, similarly to (\ref{eq:S-l-pri*})--(\ref{eq:S-l-pri}),  implies
{}
\begin{equation}
\hskip 20pt
\pm\theta +2\delta\theta_{\tt p}= -\frac{\ell}{kr}\Big(1+\frac{\ell^2}{6k^2r^2}\Big), \quad {\rm or}\quad
\ell_0 = \mp kr \sin\big(\theta\pm2\delta\theta_{\tt p}\big)+{\cal O}(\theta^5,\delta\theta_{\tt p}^3),
\label{eq:S-l-pri*p}
\end{equation}
where  $\delta\theta_{\tt p}= {d \delta^*_\ell}/{d \ell}$ is the semiclassical angle of light deflection \cite{Newton-book-2013}. This angle may be computed from (\ref{eq:delta-D*-av0WKB}) and (\ref{eq:a_b_del-r_mu0}) by accounting for the semiclassical relation $\ell=kb$. As a result, the light deflection angle is computed to be
{}
\begin{eqnarray}
\delta\theta_{\tt p}=\frac{d \delta^*_\ell}{k d b}=
\frac{\pi e^2}{m_e\omega^2}
\sum_{i\geq2} {\alpha_i}\beta_iB[{\textstyle\frac{1}{2}}\beta_i+{\textstyle\frac{1}{2}},{\textstyle\frac{1}{2}}]\Big(\frac{R_\odot}{b}\Big)^{\beta_i}.
\label{eq:ang*}
\end{eqnarray}

Using the phenomenological model (\ref{eq:model}) in (\ref{eq:ang*}), we estimate the plasma deflection angle, $\delta\theta_{\tt p}$, as a function of the impact parameter and the wavelength:
{}
\begin{eqnarray}
\delta\theta_{\tt p}&=&\Big\{
6.60\times 10^{-13}\Big(\frac{R_\odot}{b}\Big)^{16}+
2.05\times 10^{-13}\Big(\frac{R_\odot}{b}\Big)^{6}+\nonumber\\
&&\hskip 70pt
+
2.43\times 10^{-16}\Big(\frac{R_\odot}{b}\Big)^2\Big\}\Big(\frac{\lambda}{1~\mu{\rm m}}\Big)^2 ~~{\rm rad},
\label{eq:ang*ip}
\end{eqnarray}
which suggests that for sungrazing rays (i.e., for the rays with impact parameter $b\simeq R_\odot$), the bending angle (\ref{eq:ang*ip}) reaches the value of $\delta\theta_{\tt p}(R_\odot)=8.65\times 10^{-13} \,\big({\lambda}/{1~\mu{\rm m}}\big)^2$ rad, which is large for radio wavelengths, but is negligible in optical or IR bands. For typical observing situations with reasonable Sun-Earth-probe separation angles \cite{Bertotti-Giampieri:1998,DSN-handbook-2017,Verma-etal:2013}, expression (\ref{eq:ang*}) provides a good description.
Note that this expression for the plasma deflection angle, $\delta\theta_{\tt p}$, is identical to that obtained in \cite{Giampieri:1994kj,Bertotti-Giampieri:1998} and used for the recent Cassini experiment \cite{Bertotti-etal-Cassini:2003}. As the earlier result was obtained with different physical assumptions and mathematical tools, such correspondence confirms the validity of our approach, which relies on the wave-optical treatment of the problem advocating for a direct solution of Maxwell's equations.

With (\ref{eq:S-l-pri*p}), we may compute the needed expressions for the value of the phase along the path of stationary phase:
{}
\begin{eqnarray}
\varphi^{[\tt p]}_{\pm}(\ell_0)&=&
\pm{\textstyle{\frac{\pi}{4}}}+kr\big(-{\textstyle{\frac{1}{2}}}\theta^2+{\textstyle{\frac{1}{24}}}\theta^4\big)+2\delta^\star_\ell+{\cal O}(\theta^5\delta\theta_{\tt p},\delta\theta^2_{\tt p}).
\label{eq:S-l2*pd1}
\end{eqnarray}
For the second derivative of the phase along the same path, similarly to (\ref{eq:S-l2}), we have
{}
\begin{eqnarray}
\hskip -40pt
\varphi''(\ell_0)\equiv \frac{d^2\varphi^{[\tt p]}_{\pm}}{d\ell^2}\Big|_{\ell=\ell_0}&=&\frac{1}{kr}\Big(1+{\textstyle\frac{1}{2}}\theta^2+{\textstyle\frac{5}{24}}\theta^4+{\cal O}(\theta^6)\Big)+
\frac{2d^2 \delta^\star_b}{d\ell^2}+{\cal O}(\theta^6, \delta\theta^2_{\tt p}).
\label{eq:S-l2*pdd}
\end{eqnarray}

Using (\ref{eq:ang*}), we estimate the magnitude of the second term in this expression:
{}
\begin{eqnarray}
\frac{d^2 \delta^\star_b}{d\ell^2}=-\frac{1}{kb}\frac{\pi e^2}{m_e\omega^2}
\sum_{i\geq2} {\alpha_i}\beta^2_iB[{\textstyle\frac{1}{2}}\beta_i+{\textstyle\frac{1}{2}},{\textstyle\frac{1}{2}}]\Big(\frac{R_\odot}{b}\Big)^{\beta_i}.
\label{eq:ang*-dd}
\end{eqnarray}
Evaluating this quantity with the values from the empirical
model (\ref{eq:model}), we see that for the smallest impact parameter $b=R_\odot$ this quantity takes the largest value of $2{d^2 \delta^\star_b}/{d\ell^2}=1.57\times 10^{-26}\,(\lambda/(1~\mu{\rm m})^3$. This results in the fact  that for optical wavelengths, even at the heliocentric distance of $r\simeq 6.5\times 10^3$~AU, this term will be over $10^4$ times smaller than the $1/(kr)$ term in (\ref{eq:S-l2*pdd}),
representing a small correction to $\varphi''(\ell_0)$ that may be neglected for our purposes. This is equivalent to treating the deflection angle $\delta\theta_{\tt p}$ constant, which is consistent with the eikonal approximation \cite{Akhiezer-Pomeranchuk:1950,Glauber-Matthiae:1970,Semon-Taylor:1977,Sharma-etal:1988,Sharma-Somerford:1990,Sharma-Sommerford-book:2006}.

As a result, the expression for the second derivative of the phase from (\ref{eq:S-l2*pdd}) takes the form
{}
\begin{equation}
 \varphi^{[\tt p]}{}''(\ell_0)=\frac{1}{kr}\Big(1+{\textstyle\frac{1}{2}}\theta^2+{\textstyle\frac{5}{24}}\theta^4\Big)+{\cal O}\Big(\theta^6, \delta\theta_{\tt p}^2,\frac{\delta\theta_{\tt p}}{kb}\Big).
\label{eq:S-l2*p2d}
\end{equation}

The relevant, plasma-dependent part in the integral in (\ref{eq:Pi_s_exp1}) now is easy to evaluate using the method of stationary phase. Similarly to (\ref{eq:S-l3p+*}), we have the amplitude of the plasma-dependent  term in (\ref{eq:Pi_s_exp1}) evaluated to be
{}
\begin{eqnarray}
\hskip -65pt
A^{[\tt p]}(\ell_0)a(\ell_0)\sqrt{\frac{2\pi}{\varphi''(\ell_0)}}
&=&(\mp1)^{\textstyle\frac{3}{2}}k^2r^2\sqrt{1\pm\frac{\sin2\delta\theta_{\tt p}}{\sin\theta}}\sin(\theta\pm2\delta\theta_{\tt p})
\Big(1
+{\cal O}(\theta^5,\delta\theta^2_{\tt p})\Big),
\label{eq:S-l3p+**p}
\end{eqnarray}
where the superscript ${}^{[\tt p]}$ denotes the term due to the plasma phase shift. As it was done with (\ref{eq:S-l3p+*}), we can drop the $1/(ikr)$ term as it is much smaller compared to the leading term. Also, using the fact that $\sin\theta \simeq b/r$, we may evaluate the expression $({1\pm{\sin2\delta\theta_{\tt p}}/{\sin\theta}})^\frac{1}{2}\simeq 1\pm{r\delta\theta_{\tt p}}/{b}+{\cal O}(\delta\theta^2_{\tt p})$. Considering  (\ref{eq:ang*}), we see that the largest value of the bending angle, $\delta\theta_{\tt p}$,  is reached at the smallest impact parameters, $b=R_\odot$, limiting the size of this angle as  $\delta\theta_{\tt p}(R_\odot)\leq 8.65\times 10^{-13} \,\big({\lambda}/{1~\mu{\rm m}}\big)^2$ rad. As such, for optical wavelengths ${r\delta\theta_{\tt p}}/{b}$ becomes significant only beyond $10^9$~AU, which is beyond any practical significance. Therefore, we will neglect this term in our further considerations.

As a result, similarly to (\ref{eq:Pi_s_exp4+1*}), we obtain the contribution of the plasma-dependent term in  (\ref{eq:Pi_s_exp1}) in the  form
{}
\begin{eqnarray}
\delta\alpha^{[{\tt p}]}_\pm(r,\theta)&=&
E_0\sin\big(\theta\pm2\delta\theta_{\tt p}\big)
e^{i\big(kr\cos\theta+2\delta^*_\ell\big)}+{\cal O}\Big(\theta^5,\delta\theta^2_{\tt p},\frac{r\delta\theta_{\tt p}}{b}\Big).
\label{eq:Pi_s_exp4+1**p}
\end{eqnarray}

With results (\ref{eq:Pi_s_exp4+1*}) and (\ref{eq:Pi_s_exp4+1**p}) at hand, we may now present the  quantity $\alpha(r,\theta)$ from (\ref{eq:Pi_s_exp1}), with $z=r\cos\theta$,  as
{}
\begin{eqnarray}
\alpha(r,\theta)&=&\delta\alpha^{[0]}_\pm(r,\theta)+\delta\alpha^{[\tt p]}_\pm(r,\theta)=\nonumber\\
&=&
E_0\Big\{\sin\Big(\theta\pm\frac{2d\delta^\star_\ell}{d\ell}\Big)e^{i2\delta^*_\ell}-\sin\theta\Big\}e^{ikz}+
{\cal O}\Big(\theta^5,\delta\theta^2_{\tt p},\frac{r\delta\theta_{\tt p}}{b}\Big).~~~
\label{eq:Pi_s_exp4+1*=}
\end{eqnarray}

Using the approach presented above, we may now evaluate the scattering efficiency factors  $\beta(r,\theta)$ and $\gamma(r,\theta)$.

\subsection{Evaluating the scattering functions $\beta(r,\theta)$ and $\gamma(r,\theta)$}
\label{sec:amp_func}

To investigate the behavior $\beta(r, \theta)$ from (\ref{eq:beta**1}), we  need to establish the asymptotic behavior of $P^{(1)}_{l}(\cos\theta)/\sin\theta$ and $\partial P^{(1)}_{l}(\cos\theta)/\partial \theta$. For fixed $\theta$ and $\ell\rightarrow\infty$ this behavior is given\footnote{We note that, for any large $\ell$, formulae (\ref{eq:pi-l*})--(\ref{eq:tau-l*}) are insufficient in a region close to the forward $(\theta=0$) and backward ($\theta=\pi$) directions. More precisely, (\ref{eq:pi-l*})--(\ref{eq:tau-l*}) hold for $\sin\theta\gg1/\ell$ (see discussion in \cite{Turyshev-Toth:2018}.) Nevertheless, these expressions are sufficient for our purposes as in the region of interest the latter condition is satisfied.} \cite{vandeHulst-book-1981} as (this can be obtained directly from (\ref{eq:P1l<})):
{}
\begin{eqnarray}
\frac{P^{(1)}_\ell(\cos\theta)}{\sin\theta}
&=& \Big(\frac{2\ell}{\pi\sin^3\theta}\Big)^{\frac{1}{2}} \sin\Big((\ell+{\textstyle\frac{1}{2}})\theta-{\frac{\pi}{4}}\Big)+{\cal O}(\ell^{-\textstyle\frac{3}{2}}),
\label{eq:pi-l*}\\
\frac{dP^{(1)}_\ell(\cos\theta)}{d\theta}
&=&  \Big(\frac{2\ell^3}{\pi\sin\theta}\Big)^{\frac{1}{2}} \cos\Big((\ell+{\textstyle\frac{1}{2}})\theta-{\frac{\pi}{4}}\Big)+{\cal O}(\ell^{-\textstyle\frac{1}{2}}).
\label{eq:tau-l*}
\end{eqnarray}

With these approximations, the function $\beta(r,\theta)$ in the region outside the geometric shadow, takes the  form:
{}
\begin{eqnarray}
\hskip -30pt
\beta(r,\theta)&=&E_0\frac{e^{ikr}}{ikr}
\sum_{\ell=kR_\odot}^\infty \frac{(\ell+{\textstyle\frac{1}{2}})}{\ell(\ell+1)}
e^{i\big(\frac{\ell(\ell+1)}{2kr}+\frac{[\ell(\ell+1)]^2}{24k^3r^3}\big)}\Big(e^{2i\delta^*_\ell}-1\Big)
\times\nonumber\\
&&
\hskip 10pt \times\,
\Big\{
 \Big(\frac{2\ell^3}{\pi\sin\theta}\Big)^{\frac{1}{2}} \Big(1-\frac{\ell(\ell+1)}{2k^2r^2}-\frac{[\ell(\ell+1)]^2}{8k^4r^4}\Big)\cos\Big((\ell+{\textstyle\frac{1}{2}})\theta-{\frac{\pi}{4}}\Big)+\nonumber\\[3pt]
 &&\hskip 40pt +\,
 \Big(\frac{2\ell}{\pi\sin^3\theta}\Big)^{\frac{1}{2}} \sin\Big((\ell+{\textstyle\frac{1}{2}})\theta-{\frac{\pi}{4}}\Big)
\Big\}.~~~~~~~
\label{eq:S1-v0s}
\end{eqnarray}
For large $\ell\gg1$, the first term in the curly brackets in (\ref{eq:S1-v0s}) dominates, so that this expression may be given as
{}
\begin{eqnarray}
\hskip -30pt
\beta(r,\theta)&=&E_0\frac{e^{ikr}}{ikr}
\sum_{\ell=kR_\odot}^\infty
\Big(\frac{2\ell}{\pi\sin\theta}\Big)^{\frac{1}{2}} \Big(1-\frac{\ell^2}{2k^2r^2}-\frac{\ell^4}{8k^4r^4}\Big)
\Big(e^{2i\delta^*_\ell}-1\Big)\times\nonumber\\
&&\hskip 60pt \times\,
e^{i\big(\frac{\ell^2}{2kr}+\frac{\ell^4}{24k^3r^3}\big)}
\cos\Big(\ell\theta-{\frac{\pi}{4}}\Big).~~~~~~~
\label{eq:S1-v0s+}
\end{eqnarray}

To evaluate $\beta(r,\theta)$ from the expression (\ref{eq:S1-v0s+}), we again use the method of stationary phase. For this, representing (\ref{eq:S1-v0s+}) in the form of an integral over $\ell$, we have:
{}
\begin{eqnarray}
\hskip -35pt
\beta(r,\theta)&=&-E_0\frac{e^{ikr}}{kr}
\int_{\ell=kR_\odot}^\infty \hskip-4pt
\frac{\sqrt{\ell}d\ell}{\sqrt{2\pi\sin\theta}}
\Big(1-\frac{\ell^2}{2k^2r^2}-\frac{\ell^4}{8k^4r^4}\Big)
\Big(e^{2i\delta^*_\ell}-1\Big)\times\nonumber\\
&&\hskip 60pt \times\,
e^{i\big(\frac{\ell^2}{2kr}+\frac{\ell^4}{24k^3r^3}\big)}
\Big(e^{i(\ell\theta+{\textstyle\frac{\pi}{4}})}-e^{-i(\ell\theta+{\textstyle\frac{\pi}{4}})}\Big).~~~~~~~
\label{eq:S1-v0s+int*}
\end{eqnarray}

As we have done with (\ref{eq:Pi_s_exp1}), we treat this integral as a sum of  two integrals: a plasma-free and a plasma-dependent term. Expression (\ref{eq:S1-v0s+int*}) shows that the $\ell$-dependent parts of the phase have a structure identical to (\ref{eq:S-l}) and (\ref{eq:S-l*p}). Therefore, the same solutions for the points of stationary phase apply. As a result, using (\ref{eq:S-l-pri}) and (\ref{eq:S-l2}), we evaluate (\ref{eq:S1-v0s+int*}) similarly to (\ref{eq:Pi_s_exp4+1*=}) as below:
{}
\begin{eqnarray}
\beta(r,\theta)=
E_0\Big\{\cos\big(\theta\pm\frac{2d\delta^\star_\ell}{d\ell}\big)e^{i2\delta^\star_\ell}
-\cos\theta \Big\}e^{ikz}+{\cal O}(\theta^6,\delta\theta^2_{\tt p},\frac{r\delta\theta_{\tt p}}{b}).~~~
\label{eq:Pi_s_exp4+1*=*}
\end{eqnarray}

\label{sec:amp_func-der}

To determine the remaining components of the EM field (\ref{eq:DB-sol00p*}), we need to evaluate  the function $\gamma(r,\theta)$
from (\ref{eq:gamma**1}). For that, we use the asymptotic behavior of $P^{(1)}_{l}(\cos\theta)/\sin\theta$ and $\partial P^{(1)}_{l}(\cos\theta)/\partial \theta$ from (\ref{eq:pi-l*})--(\ref{eq:tau-l*}), and rely on the method of stationary phase. Similarly to (\ref{eq:S1-v0s}), we will drop the second term in the curly brackets in (\ref{eq:gamma**1}). The remaining expression for $\gamma(r, \theta)$ will now be determined by evaluating the following integral:
{}
\begin{eqnarray}
\hskip -70pt
\gamma(r, \theta)&=& -E_0\frac{e^{ikr}}{kr} \int_{\ell=kR_\odot}^\infty \hskip -3pt
\frac{\sqrt{\ell}d\ell}{\sqrt{2\pi\sin\theta}}
e^{i\big(\frac{\ell^2}{2kr}+\frac{\ell^4}{24k^3r^3}\big)}
\Big(e^{i2\delta^*_\ell}-1\Big)
\Big(e^{i(\ell\theta+{\textstyle\frac{\pi}{4}})}-e^{-i(\ell\theta+{\textstyle\frac{\pi}{4}})}\Big).~~~~~
  \label{eq:gamma**1*}
\end{eqnarray}

Clearly, this expression yields the same equation to determine the points of the stationary phase (\ref{eq:S-l}) and (\ref{eq:S-l*p}) and, thus, all the relevant results obtained in Sec.~\ref{sec:radial-comp}. As a result, we may present the integral (\ref{eq:gamma**1*}) as
{}
\begin{eqnarray}
\gamma(r,\theta)=
E_0\Big\{e^{i2\delta^*_\ell}-1\Big\}e^{ikz}+{\cal O}(\theta^6,\delta\theta^2_{\tt p},\frac{r\delta\theta_{\tt p}}{b}).
\label{eq:Pi_s_exp4+}
\end{eqnarray}

\subsection{Diffraction of light in the solar system}
\label{sec:EM-fieldsol}

At this point, we have all the necessary ingredients to present the ultimate solution for the scattered EM field in the eikonal approximation. To determine the components of the plasma-scattered EM field $({\vec E}^{\tt p},{\vec H}^{\tt p})$, we use the expressions that we obtained for the functions $\alpha(r,\theta)$, $\beta(r,\theta)$ and $\gamma(r,\theta)$, which are given by (\ref{eq:Pi_s_exp4+1*=}), (\ref{eq:Pi_s_exp4+1*=*}) and (\ref{eq:Pi_s_exp4+}), correspondingly, and substitute them in (\ref{eq:DB-sol00p*}). This allows us to compute the total field $({\vec E},{\vec H})$, which, in accord to (\ref{eq:Pi_g+p0+in*}), is given by the sum of incident EM field $({\vec E}^{\tt (0)},{\vec H}^{\tt (0)})$ from (\ref{eq:DB-t-pl=p10}) and the plasma-scattered field $({\vec E}^{\tt p},{\vec H}^{\tt p})$. The total field ${\vec E}={\vec E}^{\tt (0)}+{\vec E}^{\tt p}$ and ${\vec H}={\vec H}^{\tt (0)}+{\vec H}^{\tt p}$, up to terms of ${\cal O}(\theta^6,\delta\theta^2_{\tt p},{r\delta\theta_{\tt p}}/{b})$ is given as:
{}
\begin{eqnarray}
\hskip -60pt
  \Bigg(\hskip -4pt \begin{array}{cc}
{  \hat E}_r \\
{  \hat H}_r
  \end{array} \hskip -4pt\Bigg) &=&
  E_0
 \sin\Big(\theta\pm\frac{2d\delta^\star_\ell}{d\ell}\Big)
 \Bigg(\hskip -4pt \begin{array}{cc}
\cos\phi \\
\sin\phi  \\
  \end{array}\hskip -4pt \Bigg)e^{i\big(kz+2\delta_\ell^*-\omega t\big)},
  \label{eq:DB-t-pl=pV0}\\
  \hskip -60pt
  \Bigg(\hskip -4pt  \begin{array}{cc}
{  \hat E}_\theta \\
{  \hat H}_\theta
  \end{array} \hskip -4pt\Bigg) &=&
  E_0 \cos\theta
     \Bigg( \hskip -4pt \begin{array}{cc}
\cos\phi \\
\sin\phi  \\
  \end{array} \hskip -4pt \Bigg)
  e^{i\big(kz+2\delta^*_{\ell}-\omega t\big)},
  \label{eq:DB-t-pl=pV1}
\hskip 10pt
\Bigg( \hskip -4pt\begin{array}{cc}
{  \hat E}^{\tt }_\phi\\
{  \hat H}^{\tt }_\phi\
  \end{array} \hskip -4pt\Bigg) =
  E_0
\Bigg(\hskip -4pt \begin{array}{cc}
-\sin\phi \\
\cos\phi  \\
  \end{array} \hskip -4pt\Bigg) \,e^{i\big(kz+2\delta^*_{\ell}-\omega t\big)}.
  \label{eq:DB-t-pl=pV2}
\end{eqnarray}

This is our main result. It establishes the solution for the EM field propagating through the solar system. One can see that the total EM field behind the very large sphere, $\lambda\ll R_\odot$, embedded in the spherically symmetric plasma distribution, has the structure similar to the incident EM wave. However its phase and propagation direction are affected by the plasma. We note that plasma the defocuses light rays, introducing aberrations that are small at optical wavelengths, but are significant in the radio bands. Images taken by conventional astronomical instruments will be affected by the static plasma. Although temporal variability in the plasma may introduce additional aberrations, at optical wavelengths such effects are small and may be accounted for with standard observational techniques \cite{Huber-etal:2013,Lang-book:2009}.

\begin{figure}[t]
\hskip 1in\includegraphics{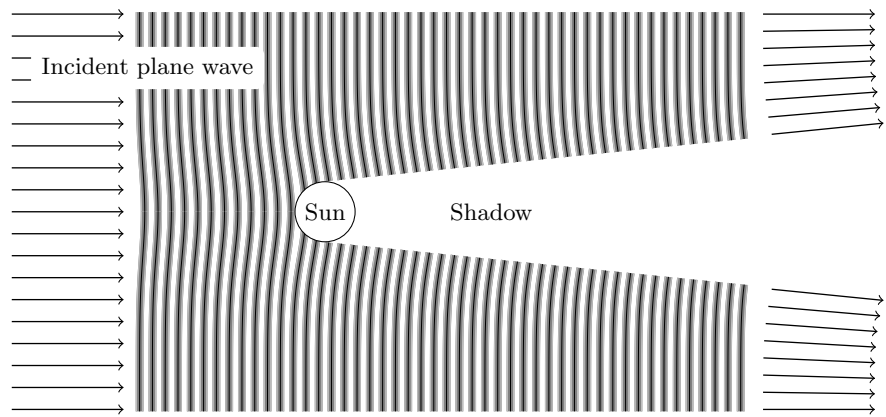}
\caption{The effects of the solar shadow and the solar atmosphere on an incident plane EM wave. (Exaggerated, not to scale.) Most of the phase shift is gained in the vicinity of the Sun, where the plasma density is dominated by
the terms with higher powers of $(R_\odot/r)$ in the empirical solar plasma model (\ref{eq:model}). Far outside the solar system, caustics form, as observed in \cite{Clegg:1997ya}.
\label{fig:caustic}}
\end{figure}

To consider the the impact of the solar plasma on the imaging properties of the astronomical telescopes, we need to know the energy flux at the image plane situated at a particular distance from the Sun, which is given by the Poynting vector, ${\vec S}$. Following the discussion in Sec.~III.F in  \cite{Turyshev-Toth:2017}, we take ${\vec E}$ and ${\vec H}$ from  (\ref{eq:DB-t-pl=pV0})--(\ref{eq:DB-t-pl=pV2}), to appropriate order, we  derive the following expression for the Poynting vector $\vec S$ in cylindrical coordinates $(r,\theta,\phi)$ \cite{Landau-Lifshitz:1988,Turyshev-Toth:2017}:
{}
\begin{eqnarray}
\hskip -30pt
\overline{{\vec S}}= \frac{c}{4\pi} \overline{
  [({\rm Re}\,{\vec E})\times ({\rm Re}\,{\vec H})]}=
\frac{c}{8\pi}  E_0^2\Big(\mp2\delta\theta_{\tt p}; \, 0; \,1\Big)+{\cal O}(\delta\theta_{\tt p}^2,\frac{r\delta\theta_{\tt p}}{b}),
  \label{eq:Sg0+}
\end{eqnarray}
where the overline, $\overline{a}$, denotes time averaging. One may see that the scattering of light in the solar corona results in the refraction of light that is characterized by the angle $\delta\theta_{\tt p}$, as expected. As shown on Fig.~\ref{fig:caustic}, the presence of plasma changes the shape of the shadow behind the Sun from a cylindrical to a conical shape, starting with a rotational hyperboloid region, with asymptotes characterized by the plasma bending angle $\delta\theta_{\tt p}$ and eventually (far outside the heliosphere at optical wavelengths) forming caustics, as it was also observed in \cite{Clegg:1997ya}.

The result given in (\ref{eq:Sg0+}) describes the angle of deflection of the light rays as they are scattered by the solar plasma, as shown in Fig.~\ref{fig:caustic}. This figure illustrates schematically that the direction of light rays changes as they traverse the solar system. The diagram also shows the surfaces of equal phase, representing the delay introduced by the plasma and the resulting tilt of the EM wavefront. The conical shape of the shadow is controlled by the ray's bending angle.

\section{Discussion and Conclusions}
\label{sec:disc}

We studied the propagation of EM waves in the vicinity of a very large opaque sphere representing our Sun, and considered the problem of diffraction of light in the presence of the extended solar corona. We considered a free electron plasma with a generic spherically symmetric and static distribution, the electron number density of which covers the entire solar system all the way to  the boundary of the heliosphere (\ref{eq:n-eps_n-ism}). We developed a wave-optical treatment of the diffraction of light in a realistic model of the plasma present in the solar system  and studied the shadow cast by a large sphere in the presence of such plasma, thereby extending the results reported in \cite{Turyshev-Toth:2018}.

In our approach we sought to solve the problem above by considering three regions that cover the solar system, namely
\begin{inparaenum}[i)]
\item the Sun itself, providing the fully absorbing boundary conditions (see Sec.~\ref{sec:bound-c}) that let us study its shadow;
\item the region within the solar system covering heliocentric distances from the solar surface to the edge of the heliosphere, $R_\odot\leq r\leq R_\star$, which yields a description of the light scattering problem that is relevant to all astronomical observations conducted within the solar system, and
\item the region that lies outside the heliosphere (see Sec.~\ref{sec:diff-large-sph}), yielding a description of light scattering in the vicinity of any star, which is relevant for modeling contributions of plasma in extrasolar systems to any type of astronomical observations.
\end{inparaenum}

Solving the wave equation for the Debye potential in the presence of an arbitrary potential is a challenging problem. In fact, no closed form analytical solutions exist. However, we were able to apply the existing tools developed to address similar problems encountered in describing nuclear and atomic scattering. One of those tools---the eikonal or high-energy approximation (see Secs.~\ref{sec:eik}--\ref{sec:eik2})---was very effective in helping us to find a solution that satisfies our objectives.   Using this method, we have shown that the presence of the plasma in the solar system results in a phase shift (\ref{eq:a_b_del-r+-tot*}) and related change in the direction of the wave's propagation (\ref{eq:ang*}), (\ref{eq:Sg0+}).

Although, our results are similar to those obtained with usual geometric optics approximation, our approach was different: We directly solved Maxwell's equations using the Mie formalism and relied on the eikonal approximation to develop the solution for the EM field from a wave-theoretical point of view. Based on the results that we obtained, we see that by tilting the wavefront, the static, spherically symmetric plasma (\ref{eq:n-eps_n-ism}) introduces optical aberrations and, thus, it affects image formation for astronomical observations
conducted in the inner solar system. Similar situations are typically encountered in many areas of practical astronomy, including
\begin{inparaenum}[i)]
\item navigation and tracking of interplanetary spacecraft where the presence of the solar plasma leads to the appearance of plasma-induced delay and is related to source brightness variations, especially for signals in the microwave frequency band (i.e., in VLBI) \cite{Giampieri:1994kj,Bertotti-Giampieri:1998,DSN-handbook-2017,Verma-etal:2013,Muhleman-Anderson:1981,Anderson-etal:2002};
\item imaging of exoplanets with the solar gravitation lens, where observations are conducted on the solar background and, thus, proper modeling of the plasma contribution is critical \cite{Turyshev-Toth:2017,Turyshev-etal:2018};
\item astrophysics research in crowded star fields where one needs to account for the plasma environment around foreground stars where the plasma's presence results in the deflection and defocusing of light and also leads to a reduction in the brightness of the observed source \cite{Clegg:1997ya,Deguchi-Watson:1987};
\item our approach may also be used to provide a wave-optical treatment for gravitational microlensing, where one typically has to deal with the plasma surrounding the lens and also leads to a reduction of the brightness of the observed source \cite{Clegg:1997ya,Deguchi-Watson:1987,Bisnovatyi-Kogan:2015dxa}; and also we emphasize that
\item physical and mathematical problems similar to those addressed in the present paper often arise in the description of nuclear and atomic scattering on various potentials, where one solves the time-independent Schr\"oedinger equation in the presence of potential tails \cite{Friedrich-book-2006,Friedrich-book-2013,Burke:2011,Glauber-Matthiae:1970,Semon-Taylor:1977}.
\end{inparaenum}

Our approach is derived from first principles and requires no {\em ad hoc} assumptions. Therefore, it may be easily augmented, e.g., by introducing latitude-dependent corrections to the plasma model and, if needed, the rotating Sun. For this, we may rely on the formalism to describe the relativistic reference frames from \cite{Turyshev-Toth:2013} that can be used to treat rotation. In addition, our rigorous treatment can also serve as a basis for further study, including reliable modeling of deviations from spherical symmetry. This work in ongoing; results will be presented elsewhere.

\section*{Acknowledgments}
This work in part was performed at the Jet Propulsion Laboratory, California Institute of Technology, under a contract with the National Aeronautics and Space Administration.

\section*{References}

\bibliographystyle{iopart-num}

\providecommand{\newblock}{}

\end{document}